\documentclass[10pt,aps,twocolumn,showpacs,superscriptaddress,floatfix,nofootinbib]{revtex4-2}

\pdfoutput=1
\usepackage[utf8]{inputenc}
\usepackage[english]{babel}
\usepackage[T1]{fontenc}
\usepackage{csquotes}
\usepackage{physics}
\usepackage{amsmath,amsfonts}

\usepackage[dvipsnames]{xcolor}
\usepackage[colorlinks=true]{hyperref}
\usepackage[capitalise]{cleveref}

\usepackage{listings}

\usepackage{multirow}
\usepackage{graphicx}

\usepackage{bm}

\begin{document}

\title{Neural Wave Functions for High-Pressure Atomic Hydrogen}

\author{David Linteau}
\email{david.linteau@epfl.ch}
\affiliation{Institute of Physics, \'{E}cole Polytechnique F\'{e}d\'{e}rale de Lausanne (EPFL), CH-1015 Lausanne, Switzerland}
\affiliation{Center for Quantum Science and Engineering, \'{E}cole Polytechnique F\'{e}d\'{e}rale de Lausanne (EPFL), CH-1015 Lausanne, Switzerland}

\author{Saverio Moroni}
\affiliation{CNR-IOM DEMOCRITOS, Istituto Officina dei Materiali and SISSA Scuola Internazionale Superiore di Studi Avanzati, Via Bonomea 265, I-34136 Trieste, Italy}

\author{Giuseppe Carleo}
\affiliation{Institute of Physics, \'{E}cole Polytechnique F\'{e}d\'{e}rale de Lausanne (EPFL), CH-1015 Lausanne, Switzerland}
\affiliation{Center for Quantum Science and Engineering, \'{E}cole Polytechnique F\'{e}d\'{e}rale de Lausanne (EPFL), CH-1015 Lausanne, Switzerland}

\author{Markus Holzmann}
\affiliation{Univ. Grenoble Alpes, CNRS, LPMMC, 38000 Grenoble, France}

\begin{abstract}
We leverage the power of neural quantum states to describe the ground state wave function of solid and liquid atomic hydrogen, including both electronic and protonic degrees of freedom. For static protons, the resulting Born-Oppenheimer energies are consistently comparable to or lower than all previous projector Monte Carlo results for systems containing up to $128$ hydrogen atoms.
The same level of accuracy is preserved upon inclusion of nuclear quantum effects, thus going beyond the Born-Oppenheimer approximation. In addition, our description overcomes major limitations of current wave functions, notably by avoiding any explicit symmetry assumption on the expected quantum crystal, and sidestepping efficiency issues of imaginary time evolution with disparate mass scales. As a first application, we examine crystal formation in an extremely high-density region up to pressure-induced melting.
\end{abstract}

\maketitle

\paragraph*{Introduction--}

Atomic and molecular hydrogen have played a fundamental role in the development of quantum mechanics and quantum chemistry.
High-pressure experiments have revealed the existence of several unexpected molecular solid phases \cite{1994_mao_high_pressure_solid_hydrogen_transition_review,2011_goncharov_vibrational_spectroscopy_hydrogen} in the long-lasting search for atomic metallic hydrogen \cite{2020_gregoryanz_hydrogen_review}.
Despite the apparent simplicity, understanding the properties of high-pressure hydrogen and deuterium remains a formidable challenge for advanced first-principle simulation methods \cite{2012_mcmahon_hydrogen_helium_extreme_conditions,2024_bonitz_hydrogen_simulation_review}.

Diffusion Monte Carlo (DMC) calculations \cite{1981_ceperley_metallic_hydrogen} have provided some of the most accurate zero-temperature descriptions. However, hydrogen simulations remain challenging for two main reasons.
First, the fixed-node approximation—inherent to DMC and required to circumvent the fermion sign problem—relies on trial wave functions of exceptional accuracy to capture the subtle physics of high-pressure phases. Second, inclusion of nuclear degrees of freedom substantially increases the computational cost: the Debye energy scale of atomic motion, much smaller than the Fermi energy scale of electrons, requires longer projection times in imaginary time evolution, and the very large proton-to-electron mass ratio ($m_p/m_e \simeq 1836$) implies much slower diffusion of the heavier particles \cite{1987_ceperley_hydrogen_high_pressures}. These difficulties are shared by related methods like reptation Monte Carlo (RMC) and finite-temperature path integral Monte Carlo (PIMC) \cite{Baroni1999,1994_pierleoni_hydrogen_plasma_pimc}.

In this work, we show that Variational Monte Carlo (VMC) with neural quantum states (NQS) can overcome these limitations by using a flexible backflow wave function, enhanced by a graph neural network that couples electrons and protons.
This approach naturally incorporates nuclear quantum effects beyond the Born–Oppenheimer approximation (BOA), a regime where conventional methods struggle due to mass-scale disparities.

Furthermore, our wave function  avoids ad hoc symmetry constraints, a crucial feature for exploring high-pressure hydrogen phases where crystalline structures are unknown. As a first application, we study the onset of crystal formation in an ultra-dense region where pressure-induced melting is expected.

Since the first quantum Monte Carlo (QMC) studies on hydrogen \cite{1981_ceperley_metallic_hydrogen,1987_ceperley_hydrogen_high_pressures}, only a few works have treated electronic and protonic degrees of freedom on equal footing \cite{1993_natoli_solid_hydrogen,Natoli95,2005_holzmann_hydrogen}. Advances in QMC stem from decoupling electronic and protonic motions within the BOA. This underlies coupled electron ion Monte Carlo (CEIMC) \cite{2004_pierleoni_ceimc_hydrogen} and ab-initio molecular dynamics (MD) \cite{2008_attaccalite_md_hydrogen} methods, operating at temperatures
 substantially below the electronic Fermi temperature, but high enough to describe nuclear motion within the adiabatic approximation (i.e. the BOA).
Within the BOA, robust QMC trial wave functions are built from Hartree-Fock or density functional theory (DFT) band structure methods \cite{Wang90,1993_natoli_solid_hydrogen,Natoli95,Pierleoni08}, and improved with backflow coordinates \cite{2003_holzmann_backflow_electron_gas_and_hydrogen}.
The nuclear motion is then obtained via MD or PIMC \cite{Pierleoni2006,2012_mcmahon_hydrogen_helium_extreme_conditions,Holzmann24}, or,
more recently, using deep network free energy methods \cite{Xie23,Dong25}.

However, so far, CEIMC or MD simulations have been performed based on VMC calculations of the BOA energy surface. Although changes due to more accurate RMC or DMC values can be estimated a posteriori via reweighting of selected trajectories \cite{Ruggeri2020}, or based on machine-learned force fields trained on DMC calculations of selected proton configurations \cite{Tirelli2022,2023_niu_qmc_database_hydrogen,goswami2024high,Ly2025},
the accuracy of the standard trial wave functions employed remains a major bottleneck of these methods.
Therefore, we first benchmark our NQS on static proton configurations obtaining BOA energies that reach or surpass previous DMC/RMC values. This demonstrates the flexibility of backflow-based NQS for high-pressure hydrogen, as seen in studies of other systems such as atoms and molecules \cite{2019_holzmann_backflow,2020_pfau_deep_net,2020_hermann_neural_net_electrons,2023_gao_generalizing_neural_wavefunctions,Gao24,2024_scherbela_transferable_fermionic_neural_wf,Zhang25,hermann_ab_2023,scherbela_accurate_2025}, as well as electron matter \cite{Holzmann20,Wilson23,LiLi,Cassella23,Smith24,luo_simulating_2024,luo_solving_2025}.

We then address the ground state of the full Schrödinger equation of the combined electron-proton system—the “dynamic” case—thereby going beyond the BOA. Unlike BOA wave functions with orbitals tied to a fixed proton configuration, our NQS explicitly incorporates proton motion via a factor capturing zero-point motion.
Whereas projection methods, such as DMC and RMC, are burdened by mass-dependent convergence issues (as the kinetic energy of the propagator
ties the diffusion of the particles to their inverse mass), in VMC the convergence does not depend on the nuclear mass, because Monte Carlo updates can be addressed on the appropriate scale for each type of particle.
Finally, our ansatz accurately describes liquid and solid phases in a translationally invariant form, without imposing a particular crystal symmetry \cite{2018_ruggeri_backflow,2024_linteau_helium_4_2d}, which is crucial for high-pressure hydrogen where the phase diagram is not fully determined. We present preliminary results in the region where atomic hydrogen is expected to melt at ultra-high pressures.

\paragraph*{Method--}
We consider $N$ hydrogen atoms in a simulation cell of volume
$V = 4 \pi N r_s^3 a_0^3/ 3$, where $r_s = a / a_0$ parametrizes the electronic number density, with $a$ and $a_0$ being the mean electronic distance and the Bohr radius, respectively.
To simplify the notation, we will focus on the case of a cubic box of extension $L=V^{1/3}$ in the following.
We denote the electron and proton coordinates by $\{\mathbf{r}_i\}$ and $\{\mathbf{R}_I\}$, respectively.
As we simultaneously sample both sets of coordinates using Monte Carlo, a configuration is denoted $\mathbf{X} \equiv \{\mathbf{r},\mathbf{R}\}\equiv \{\mathbf{r}_1,\hdots, \mathbf{r}_N, \mathbf{R}_1, \hdots, \mathbf{R}_N\}$.
We periodize the simulation cell so that the Hamiltonian in atomic units reads
\begin{equation} \label{eq:hamiltonian}
    H = -\sum_{a=1}^{2N} \frac{1}{2 m_a} \nabla_{\mathbf{x}_a}^2 + \frac{1}{2} \sum_\mathbf{n \in \mathbb{Z}^3} \sum_{a,b=1}^{2N} {}^{\hspace{-1mm}'} \frac{q_a q_b}{|\mathbf{x}_a - \mathbf{x}_b + \mathbf{n} L|},
\end{equation}
where  $\mathbf{x}_a \in \mathbf{X}$, $m_a$ and $q_a$ are respectively the mass and electric charge of particle $a$.
The restriction on the Coulomb sum, denoted by an apostrophe, specifies that $\mathbf{n} = \mathbf{0}$ is omitted when $a = b$.
The Ewald procedure \cite{1921_ewald} is used to evaluate the (conditionally convergent) second term in \cref{eq:hamiltonian}.

To construct a ground state trial wave function, it is convenient to consider the following general form
\begin{equation} \label{eq:trial_wave function}
    \Psi(\mathbf{X}) = \Phi(\mathbf{R})  \mathrm{det}[\phi_k(\mathbf{r}_i|\mathbf{X})]
    e^{-U(\mathbf{X})},
\end{equation}
where $U(\mathbf{X})$ denotes a symmetric correlation factor, $\Phi(\mathbf{R})$ is a pure nuclear wave function, and $\{\phi_k(\mathbf{r}_i|\mathbf{X})\}$ is a set of electronic orbitals, with $1 \le k \le N$, which may symmetrically depend on all coordinates in $\mathbf{X}$ that are different from $\mathbf{r}_i$.
Although this form is general for all fermionic ground state wave functions \cite{2015_taddei_iterative_backflow_fermi_liquids,Ceperley91,Clark19}, there is no guarantee that the orbitals $\phi_k(\mathbf{r}_i|\mathbf{X})$
can be efficiently represented numerically.

In the first calculations on dense hydrogen \cite{1981_ceperley_metallic_hydrogen,1987_ceperley_hydrogen_high_pressures},
$U(\mathbf{X})$ was chosen as a sum of pairwise correlation factors in terms of electron-electron (e-e), electron-proton (e-p) and proton-proton (p-p) pseudo-potentials.
The orbital part of the wave function consisted of Slater determinants for spin-up and spin-down electrons occupying either plane wave orbitals with wave vectors $\mathbf{k}$ inside the Fermi surface, or Gaussians localized at the crystalline lattice sites, denoted $\{\mathbf{R}_I^{(0)}\}$.
Later calculations \cite{Natoli95} have used orbitals from DFT within the local-density approximation (LDA), with the orbitals also pinned at $\{\mathbf{R}_I^{(0)}\}$.
Exchange effects for protons (and deuterium) are expected to be small in the crystalline phase \cite{1981_ceperley_metallic_hydrogen}.
Quantum statistical effects on protons have thus been neglected, and the proton crystal was described by the asymmetric (Nosanow) wave function   $\Phi(\mathbf{R}) = \prod_I \varphi(|\mathbf{R}_I - \mathbf{R}_I^{(0)}|)$ \cite{1964_nosanow_crystalline_he3,1968_hansen_gs_he4_he3}, where $\varphi$ is a Gaussian function with a width variational parameter to optimize.
Although DMC stochastically improves the wave function via imaginary time projection, in practice the results are strongly biased by the explicit dependence on the assumed crystalline structure. Full exploration of the nuclear phase space in imaginary time is difficult due to the large mass imbalance between protons and electrons.
We note that the effect is amplified when considering deuterium having twice the proton's mass, obeying bosonic nuclear statistics.
Backflow and three-body correlations in hydrogen \cite{2003_holzmann_backflow_electron_gas_and_hydrogen} have shown to improve energies also for the fully dynamic protons at zero-temperature  \cite{2005_holzmann_hydrogen}, but have not been explored further  since.

Our wave function builds upon these previous ones by incorporating two important modifications.
First, as electronic orbitals we use
\begin{equation} \label{eq:tabc_orbitals_static}
    \phi_\mathbf{k}^\sigma(\mathbf{r}_i) = \sum_{\mathbf{n} \in \mathbb{Z}^3} \sum_{I=1}^N e^{i \mathbf{k} \cdot (\mathbf{R}_I + \mathbf{n} L)} \chi(\mathbf{r}_i^\sigma - \mathbf{R}_I -  \mathbf{n} L),
\end{equation}
which coincide with Bloch functions for static protons on a crystalline lattice.
For $\chi$, we have chosen a Gaussian with a free width parameter.
While these radially symmetric s-orbitals seem to be sufficient at describing the range of densities studied in this work, angular dependence via higher orbitals (p, d, f, etc.) can be included in a straightforward manner.

Second, we increase the expressivity of the trial wave function by adding propagator-like backflow to the electron coordinates,
\begin{equation} \label{eq:electronic_backflow}
    \mathbf{r}_i \to \mathbf{r}_i + W \mathbf{y}_i^{(b)},
\end{equation}
where $\mathbf{y}_i^{(b)}$ is the electron vertex output of a message-passing neural network (MPNN) after $b$ iterations, and $W$ is a matrix of complex variational parameters.
This electron backflow is only added in the orbital part of the wave function.
For $U(\mathbf{X})$, we take advantage of the electron edge output of the MPNN, denoted by $\mathbf{Y}_{ij}^{(b)}$, to create a propagator-like term, similarly as in Ref.~\cite{2024_linteau_helium_4_2d}.
With this additional piece the correlation factor $U(\mathbf{X})$ reads
\begin{equation} \label{eq:edge_jastrow_electron_part}
    U(\mathbf{x}) \to U(\mathbf{x}) + \sum_{i<j} \left[w \ \mathrm{MLP}(\mathbf{Y}_{ij}^{(b)}) \right],
\end{equation}
where MLP stands for multilayer perceptron and $w \in \mathbb{R}$ is a variational parameter.
The MPNN operates on two underlying graphs: the electron one and the proton one, which interact at each message-passing iteration.
In particular, it involves e-e and e-p contributions, meaning that the resulting backflow involve both the electron and proton coordinates, with the e-p contributions capturing a smaller—though important—part of the correlation
(see \cref{appendix:mpnn_implementation}).

This corresponds to a modified version of the message-passing neural quantum state architecture introduced in Ref.~\cite{2024_pescia_mpnn_electrongas}.
In the dynamic case, when the protons are not localized to lattice sites, the zero-point motion of the protons is captured with the edge output $\mathbf{Y}_{IJ}^{(b)}$ of a different MPNN than the electronic one, which is processed similarly as in \cref{eq:edge_jastrow_electron_part}.
This latter MPNN operates only on a single graph—the proton one.

Although we can reach large system sizes of around $10^2$
particles with our implementation (see the Results section), important finite-size effects remain, which can be addressed from strategies discussed in Ref.~\cite{2016_holzmann_finite_size_effects}.
Most important, especially for metallic systems, are shell effects due to the sharp Fermi surface.
These shell effects can however be strongly reduced by employing twist-averaged boundary conditions (TABC) \cite{2001_lin_ceperley_tabc}, which, for a displacement of particle $i$ by one lattice vector $L \mathbf{e}_\alpha$ along direction $\alpha$, impose the following constraint on the wave function

\begin{equation} \label{eq:tabc_constraint}
  \Psi(\{\mathbf{r}_j + \delta_{ij}L\mathbf{e}_\alpha\},\mathbf{R}) = e^{i\vartheta_\alpha}\Psi(\{\mathbf{r}_j\},\mathbf{R}),
\end{equation}
where $\mathbf{e}_\alpha$ denotes the unit vector in the direction of $\alpha \in \{x,y,z\}$, with $-\pi \le \vartheta_\alpha < \pi$, and $i,j$ labels any of the
electron coordinates.
We remark that we have only imposed TABC on the electronic degrees of freedom.
Periodic boundary conditions (PBC) correspond to the special case
where the twist angle is trivial, that is $\boldsymbol{\vartheta} \equiv (\vartheta_x,\vartheta_y,\vartheta_z)
= \boldsymbol{0}$.
We can impose TABC on the wave function by selecting the ``twisted'' wave vector $\mathbf{k} = (2 \pi \mathbf{m} + \boldsymbol{\vartheta})/L$, with $\mathbf{m} \in \mathbb{Z}^3$, for the electronic orbitals.
In practice, the electron orbitals are then filled by taking the twisted wave vectors with the smallest norm.

\paragraph*{Results--}

\begin{table}[b]
    \centering
    \begin{tabular}{c|c|c|c}
    $N$ & Wave function & $E/N$ & $\sigma^2/N$ \\
    \hline \hline
    \multirow{4}{*}{16} & SJ-PW (DMC)& -0.4857(1) & 0.0773(25) \\
    & SJ-LDA (DMC) & -0.4890(5) & - \\
    & BF-PW (DMC) & -0.4905(1) & 0.0232(1) \\
    & \textbf{NQS (VMC)} & \textbf{-0.49154(1)} & \textbf{0.0062(1)} \\
    \hline
    \multirow{5}{*}{54} & SJ-PW (DMC) & -0.5329(1) & 0.0642(9) \\
    &  BF-PW (DMC) & -0.5382(1) & 0.0222(2) \\
    & SJ-LDA (DMC) & -0.5390(5) & - \\
    & BF-PBE (RMC) & -0.54009(2) & 0.00565(1)
    \\
    & \textbf{NQS (VMC)} & \textbf{-0.54009(2)} & \textbf{0.00461(3)} \\
    \hline
    \multirow{5}{*}{128} & SJ-PW (DMC) & -0.4900(2) & 0.0656(23) \\
    & BF-PW (DMC) & -0.4978(4) & 0.030(1) \\
    & SJ-LDA (DMC) & -0.4978(2) & - \\
    &  BF-PBE (RMC) & -0.49922(2) & 0.00666(2)
    \\
    & \textbf{NQS (VMC)} & \textbf{-0.49962(2)} & \textbf{0.00721(2)} \\
    \hline
    \end{tabular}
    \caption{
    Static hydrogen with protons pinned to a BCC lattice at $r_s=1.31$ using PBC.
    The energy per atom, $E/N$, is given in units of Hartree, and $\sigma^2$ denotes the energy variance.
    The NQS energy, obtained from VMC calculations using the ansatz given in \cref{eq:trial_wave function}, is compared to previous DMC reference energies taken from Ref.~\cite{2003_holzmann_backflow_electron_gas_and_hydrogen}, based on Slater-Jastrow (SJ-PW) and backflow (BF-PW) plane wave orbitals, as well as Slater-Jastrow calculations using DFT-LDA (SJ-LDA) orbitals, and backflow PBE-DFT (BF-PBE) orbitals.
    For DMC results, we report the energy variance $\sigma^2$ of the underlying VMC wave function.
    Boldface entries indicate the lowest energy value within each set of compared results.
    }
    \label{table:static_hydrogen}
\end{table}

As a first step, we want to establish the accuracy of our NQS compared to previous VMC and DMC results.
Prior calculations have almost exclusively focused on BOA ground state of electrons using the external (Coulomb) potential of the static protons.
Formally, the electronic Hamiltonian, $H_{e}(\mathbf{R})$, corresponds to infinite proton mass in Eq.~(\ref{eq:hamiltonian}),

\begin{equation} \label{eq:HBO}
    H_e(\mathbf{R}) = -\frac{1}{2 m_e} \sum_{i=1}^{N} \nabla_{\mathbf{r}_i}^2 + \frac{1}{2} \sum_\mathbf{n \in \mathbb{Z}^3} \sum_{a,b=1}^{2N} {}^{\hspace{-1mm}'} \frac{q_a q_b}{|\mathbf{x}_a - \mathbf{x}_b + \mathbf{n} L|}.
\end{equation}
To obtain upper bounds to the BOA ground state energies,
we minimize $H_e(\mathbf{R})$ over electronic wave functions
$\Psi_e(\mathbf{r}|\mathbf{R})$ with purely parametric dependence on $\mathbf{R}$.
We have used the same functional form as in \cref{eq:trial_wave function}, simply treating the protonic coordinates as (external) static parameters.
Additional implicit parametric dependence on $\mathbf{R}$ may be introduced via the optimization parameters.

In \cref{table:static_hydrogen}, we compare our results for static protons localized on a perfect body-centered cubic (BCC) lattice around metallization, at $r_s=1.31$, with DMC calculations using Slater-Jastrow (SJ) and backflow wave functions with metallic plane wave orbitals \cite{2003_holzmann_backflow_electron_gas_and_hydrogen}, as well as with LDA-DFT orbitals \cite{Natoli95} and backflow augmented PBE-DFT orbitals (BF-PBE) \cite{Morales14}.
Calculations have been performed for $N=16$, $54$, and $128$ electrons under PBC.
Our VMC results systematically lower the energies per atom
by $\sim 1$ mHa compared to the best SJ-DMC calculations reported in the literature,
and perform similarly to RMC results with BF-PBE wave functions.
The quality of the wave function is further quantified by roughly a fivefold reduction in variance compared to the SJ-VMC wave function underlying the DMC calculation.
Notably, about the same improvement is obtained for all three system sizes
using a total of 2889 parameters, illustrating the size consistency of our wave function and of the results. \

In \cref{table:static_hydrogen_rs=1_rs=1.4_pbc_and_tabc} (see \cref{appendix:energy_comparison}), we further provide explicit comparisons of twist-averaged calculations for static protons in different crystal structures obtained with RMC in Ref.~\cite{Pierleoni08},
which also provide DFT orbitals augmented with backflow wave functions, as used in CEIMC calculations. Again, our NQS is able to compete with the best projector Monte Carlo results,
in this case RMC based on LDA or optimized independent
particle (IPP) orbitals augmented with backflow.

So far, we have shown that our NQS wave function reaches and improves on previous DMC energies from the literature by focusing on high-symmetry configurations of atomic hydrogen.
Our NQS state can therefore serve to enhance the accuracy of machine-learned effective potentials for hydrogen atoms \cite{zong2020understanding,Cheng2020,karasiev2021liquid,Tirelli2022,2023_niu_qmc_database_hydrogen,goswami2024high,istas2024liquidliquidphasetransitionhydrogen,tenti2025hydrogen,Ly2025}, to benchmark DFT functionals on representative proton configurations in the atomic regime of the high-pressure phase diagram \cite{Clay2016,cozza2025}, and to further achieve DMC/RMC-level precision of the BOA energy surface in future CEIMC or MD calculations.

We have further explored the expressivity of our ansatz in the molecular phases at lower pressures. Comparing to generic configurations with $N=96$ protons provided in the dense hydrogen DMC database from Ref.~\cite{2023_niu_qmc_database_hydrogen,QMC-hamm}, spanning molecular solid and liquid configurations at pressures from 50 to 200 GPa, our present NQS architecture remains roughly 2-5 mHa above the DMC reference energies, indicating limitations of our orbitals in the molecular phases.
Further work is needed to reduce the systematic bias of our ansatz in the molecular region to simultaneously capture both molecular and atomic phases using a universal ansatz, similar to recent foundation models \cite{2025_rende_foundation_neural_network_quantum_states,2024_scherbela_transferable_fermionic_neural_wf,Gao24}.
Details on the global optimization scheme that can be used for such universal ans\"atze, as well as preliminary results, are provided in \cref{appendix:global_optimization}.

\begin{table}[t]
    \centering
    \begin{tabular}{c|c|c|c}
    $N$ & Wave function & $E/N$ & $\sigma^2/N$ \\
    \hline \hline
    \multirow{4}{*}{16} & SJ-LDA (VMC) & -0.46785(2) & -\\
    &BF-PW (VMC) & -0.4724(1) & 0.030(2)  \\
    & BF-PW (DMC) & -0.4792(1) & - \\
    & \textbf{NQS (VMC)} & \textbf{-0.48091(5)} & \textbf{0.0114(1)} \\
    \hline
    \multirow{5}{*}{54} &  SJ-LDA (VMC) & -0.5195(2) & - \\
    & SJ-LDA (DMC) & -0.52415(5) & - \\
    & BF-PW (VMC) & -0.52194(5) & 0.025(1) \\
    & BF-PW (DMC) & -0.52610(7) & - \\
    & \textbf{NQS (VMC)} & \textbf{-0.52854(9)} & \textbf{0.0088(1)} \\
    \hline
    \end{tabular}
    \caption{
    Ground state energies for dynamic hydrogen at $r_s=1.31$, for different wave functions obeying PBC.
    The reference VMC and DMC energies for Slater-Jastrow with LDA-DFT orbitals (SJ-LDA) and backflow plane wave orbitals (BF-PW) are taken from Ref.~\cite{2005_holzmann_hydrogen}. For all calculations, the nuclear part of the trial wave function is given by Gaussians localized around BCC lattice sites.
    Boldface entries indicate the lowest energy value within each set of compared results.
    }
    \label{table:dynamic_hydrogen}
\end{table}

\begin{figure}[b]
    \centering
    \includegraphics[width=\columnwidth]{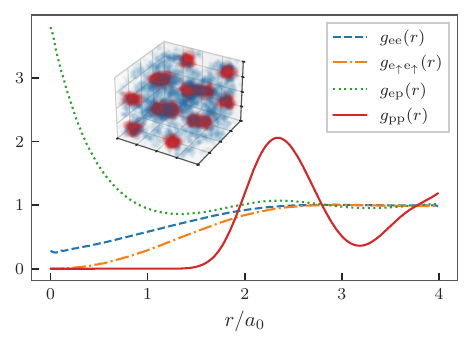}
    \caption{
    Four pair correlation functions are shown for $N=54$ between three different types of particles: ``e'', ``p'' and ``e$_{\uparrow}$'', corresponding to electrons, protons and spin-up electrons respectively.
    An inset with 300 Monte Carlo configurations in the BCC crystal is depicted (for $N=16$, for clarity), where protons form the localized red balls while electrons form the delocalized blue cloud.
    }
    \label{fig:pair_correlation_function_and_bcc_crystal}
\end{figure}

\begin{figure*}
    \centering
    \includegraphics[width=0.95\textwidth]{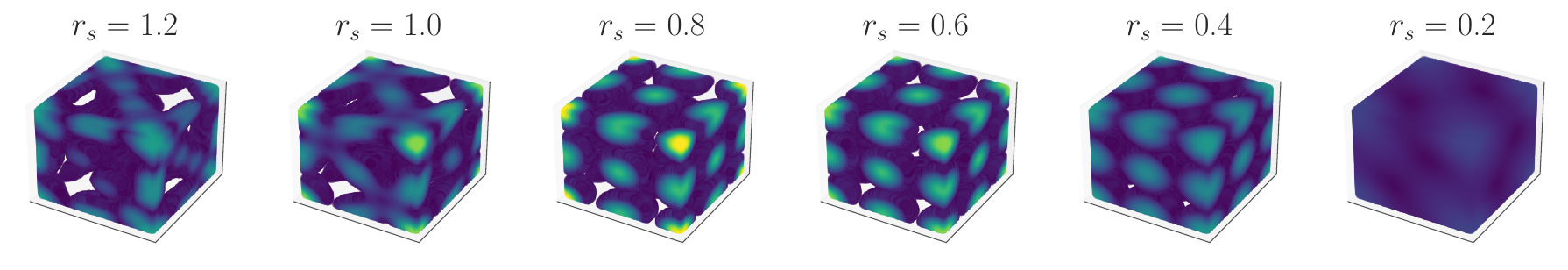}
    \caption{
    Pair correlation function for $N=8$ hydrogen atoms at different densities under PBC.
    The same color scale is used in all subplots, globally adjusted with a power law normalization to enhance visibility of the structural features.
    A fixed threshold is manually set to prevent plotting areas where the pair correlation function has low values.
    At $r_s=0.2$, the structure is drastically reduced suggesting that the system turned liquid.
    }
    \label{fig:pair_corerlation_function_panel_plot_N=8}
\end{figure*}

We finally present the calculations of ground state energies with dynamical protons without relying on the BOA. There, electronic and protonic degrees of freedom are sampled according to the full wave function, given in \cref{eq:trial_wave function}, and energies are obtained from minimizing the expectation value of $H$, given in \cref{eq:hamiltonian}. Effects beyond BOA are naturally included in the evaluation of the protons' kinetic energy, acting on the purely nuclear part $\Phi(\mathbf{R})$, as well as on the electronic wave function $\Psi_e(\mathbf{r}|\mathbf{R})$.
As a first benchmark, we consider hydrogen at $r_s=1.31$, with localized Gaussians for the nuclear wave function $\Phi(\mathbf{R})$, so that the proton positions fluctuate around BCC lattice sites.
This allows us to faithfully compare in \cref{table:dynamic_hydrogen} with VMC and DMC results from Ref.~\cite{2005_holzmann_hydrogen}, based on BF-PW and SJ-LDA trial wave functions.
Again, improvement is consistently obtained.
To provide more insights into the structure of the optimized wave function, we further show in \cref{fig:pair_correlation_function_and_bcc_crystal} pair correlation functions between different particle types (electrons, protons and spin-up electrons), as well as Monte Carlo configurations in position space showcasing the BCC lattice structure.

We finally fully relax the nuclear wave function—keeping the wave function fully translationally invariant—and explore the ground state structures of $N=8$ atoms under increasing compression.
The associated nuclear wave function $\Phi(\mathbf{R})$ is  constructed with $\mathbf{Y}_{IJ}^{(b)}$, as described above, as well as a p-p pseudo-potential.
Importantly, it does not involve a product of Gaussian functions localizing the protons at lattice sites.
\cref{fig:pair_corerlation_function_panel_plot_N=8} shows the (proton-proton) pair correlation function in three dimensional space, at various pressures near the expected melting transition.
The method used is clearly able to find structural changes when increasing the pressure (or lowering $r_s$) and the system turns liquid around $r_s \approx 0.2$.
However, studies at larger system sizes and different cell geometries are required to precisely determine the crystalline structures as well as the exact melting density.
So far, pressure-induced melting of hydrogen has only been studied from approximate matching of hydrogen to calculations for the one-component plasma \cite{Jones96}, or based on an approximate Thomas-Fermi screening model that uses a Yukawa pair-potential for the proton-proton interaction \cite{1980_mon_high_pressure_hydrogen_melting,1989_ceperley_qmc_high_pressure_melting,2006_militzer_melting_transition_hydrogen}.

The NQS-based approach presented here enables the direct identification and characterization of atomic crystalline hydrogen structures and their melting transition, and is only limited in this study by finite-size effects.

Although melting occurs at ultra-high densities, the inter-protonic distance $a^p$ remains larger than the protonic Bohr radius ($a_0^p\equiv a_0 m_e/m_p$), that is, $a^p/a_0^p \simeq 1836 \, r_s \gg 1$.
This justifies neglecting fermionic statistics in the solid phase, which means in practice that no protonic determinants are included in the wave function.
Finite-size effects, by imposing TABC for instance, as well as Fermi statistics for protons and isotope effects, will be addressed in future work.

\paragraph*{Conclusion--}
We have introduced a neural quantum state that accurately describes electron-ion systems. Validated against RMC and DMC results, our method achieves high precision for high-pressure atomic hydrogen both in the Born–Oppenheimer regime and for the full zero–temperature ground state beyond BOA. Our preliminary study demonstrates that our NQS can capture the zero–temperature melting of atomic hydrogen while treating protons and electrons on equal footing. These attributes pave the way for an extensive exploration of the high–pressure phase diagram for both hydrogen and deuterium. Moreover, the incorporation of nuclear quantum statistics opens promising avenues for studying isotope effects and nuclear spin phenomena such as the ortho–para transitions in molecular hydrogen.

The simulations were performed using NetKet \cite{2019_netket,2022_netket_3}, built on JAX \cite{2018_jax} and mpi4jax \cite{2021_hafner_vicentini_mpi4jax}. 
The data underlying the figures, together with the code used to generate them, are publicly available in the GitHub repository \cite{github_repo}.
The Laplacian was evaluated using the forward Laplacian method \cite{2024_li_forward_laplacian}, implemented via either folx \cite{2023_gao_folx} or fwdlap \cite{2024_chen_fwdlap}.

\paragraph*{Data availability.} 
The data that supports the findings of this article are openly available \cite{github_repo}.

\paragraph*{Acknowledgments--}
We thank Y. Yang for providing help with the Dense hydrogen DMC database \cite{2023_niu_qmc_database_hydrogen,QMC-hamm}, C. Pierleoni for providing BF-PBE results, and acknowledge useful discussions with G. Mazzola.
The authors acknowledge support from SEFRI under Grant No.\ MB22.00051 (NEQS - Neural Quantum Simulation).
This work was also supported by a grant from the Swiss National Supercomputing Centre (CSCS) under project ID a117 on Alps, as part of the Swiss AI Initiative.

\clearpage
\twocolumngrid
\bibliography{main.bib}

\begin{thebibliography}{82}%
\makeatletter
\providecommand \@ifxundefined [1]{%
 \@ifx{#1\undefined}
}%
\providecommand \@ifnum [1]{%
 \ifnum #1\expandafter \@firstoftwo
 \else \expandafter \@secondoftwo
 \fi
}%
\providecommand \@ifx [1]{%
 \ifx #1\expandafter \@firstoftwo
 \else \expandafter \@secondoftwo
 \fi
}%
\providecommand \natexlab [1]{#1}%
\providecommand \enquote  [1]{``#1''}%
\providecommand \bibnamefont  [1]{#1}%
\providecommand \bibfnamefont [1]{#1}%
\providecommand \citenamefont [1]{#1}%
\providecommand \href@noop [0]{\@secondoftwo}%
\providecommand \href [0]{\begingroup \@sanitize@url \@href}%
\providecommand \@href[1]{\@@startlink{#1}\@@href}%
\providecommand \@@href[1]{\endgroup#1\@@endlink}%
\providecommand \@sanitize@url [0]{\catcode `\\12\catcode `\$12\catcode
  `\&12\catcode `\#12\catcode `\^12\catcode `\_12\catcode `\%12\relax}%
\providecommand \@@startlink[1]{}%
\providecommand \@@endlink[0]{}%
\providecommand \url  [0]{\begingroup\@sanitize@url \@url }%
\providecommand \@url [1]{\endgroup\@href {#1}{\urlprefix }}%
\providecommand \urlprefix  [0]{URL }%
\providecommand \Eprint [0]{\href }%
\providecommand \doibase [0]{https://doi.org/}%
\providecommand \selectlanguage [0]{\@gobble}%
\providecommand \bibinfo  [0]{\@secondoftwo}%
\providecommand \bibfield  [0]{\@secondoftwo}%
\providecommand \translation [1]{[#1]}%
\providecommand \BibitemOpen [0]{}%
\providecommand \bibitemStop [0]{}%
\providecommand \bibitemNoStop [0]{.\EOS\space}%
\providecommand \EOS [0]{\spacefactor3000\relax}%
\providecommand \BibitemShut  [1]{\csname bibitem#1\endcsname}%
\let\auto@bib@innerbib\@empty
\bibitem [{\citenamefont {Mao}\ and\ \citenamefont
  {Hemley}(1994)}]{1994_mao_high_pressure_solid_hydrogen_transition_review}%
  \BibitemOpen
  \bibfield  {author} {\bibinfo {author} {\bibfnamefont {H.-k.}\ \bibnamefont
  {Mao}}\ and\ \bibinfo {author} {\bibfnamefont {R.~J.}\ \bibnamefont
  {Hemley}},\ }\bibfield  {title} {\bibinfo {title} {Ultrahigh-pressure
  transitions in solid hydrogen},\ }\href
  {https://doi.org/10.1103/RevModPhys.66.671} {\bibfield  {journal} {\bibinfo
  {journal} {Rev. Mod. Phys.}\ }\textbf {\bibinfo {volume} {66}},\ \bibinfo
  {pages} {671} (\bibinfo {year} {1994})}\BibitemShut {NoStop}%
\bibitem [{\citenamefont {Goncharov}\ \emph {et~al.}(2011)\citenamefont
  {Goncharov}, \citenamefont {Hemley},\ and\ \citenamefont
  {Mao}}]{2011_goncharov_vibrational_spectroscopy_hydrogen}%
  \BibitemOpen
  \bibfield  {author} {\bibinfo {author} {\bibfnamefont {A.~F.}\ \bibnamefont
  {Goncharov}}, \bibinfo {author} {\bibfnamefont {R.~J.}\ \bibnamefont
  {Hemley}},\ and\ \bibinfo {author} {\bibfnamefont {H.-k.}\ \bibnamefont
  {Mao}},\ }\bibfield  {title} {\bibinfo {title} {Vibron frequencies of solid
  {H2} and {D2} to 200 {GPa} and implications for the {P–T} phase diagram},\
  }\href {https://doi.org/10.1063/1.3574009} {\bibfield  {journal} {\bibinfo
  {journal} {The Journal of Chemical Physics}\ }\textbf {\bibinfo {volume}
  {134}},\ \bibinfo {pages} {174501} (\bibinfo {year} {2011})}\BibitemShut
  {NoStop}%
\bibitem [{\citenamefont {Gregoryanz}\ \emph {et~al.}(2020)\citenamefont
  {Gregoryanz}, \citenamefont {Ji}, \citenamefont {Dalladay-Simpson},
  \citenamefont {Li}, \citenamefont {Howie},\ and\ \citenamefont
  {Mao}}]{2020_gregoryanz_hydrogen_review}%
  \BibitemOpen
  \bibfield  {author} {\bibinfo {author} {\bibfnamefont {E.}~\bibnamefont
  {Gregoryanz}}, \bibinfo {author} {\bibfnamefont {C.}~\bibnamefont {Ji}},
  \bibinfo {author} {\bibfnamefont {P.}~\bibnamefont {Dalladay-Simpson}},
  \bibinfo {author} {\bibfnamefont {B.}~\bibnamefont {Li}}, \bibinfo {author}
  {\bibfnamefont {R.~T.}\ \bibnamefont {Howie}},\ and\ \bibinfo {author}
  {\bibfnamefont {H.-K.}\ \bibnamefont {Mao}},\ }\bibfield  {title} {\bibinfo
  {title} {{Everything you always wanted to know about metallic hydrogen but
  were afraid to ask}},\ }\href {https://doi.org/10.1063/5.0002104} {\bibfield
  {journal} {\bibinfo  {journal} {Matter and Radiation at Extremes}\ }\textbf
  {\bibinfo {volume} {5}},\ \bibinfo {pages} {038101} (\bibinfo {year}
  {2020})}\BibitemShut {NoStop}%
\bibitem [{\citenamefont {McMahon}\ \emph {et~al.}(2012)\citenamefont
  {McMahon}, \citenamefont {Morales}, \citenamefont {Pierleoni},\ and\
  \citenamefont {Ceperley}}]{2012_mcmahon_hydrogen_helium_extreme_conditions}%
  \BibitemOpen
  \bibfield  {author} {\bibinfo {author} {\bibfnamefont {J.~M.}\ \bibnamefont
  {McMahon}}, \bibinfo {author} {\bibfnamefont {M.~A.}\ \bibnamefont
  {Morales}}, \bibinfo {author} {\bibfnamefont {C.}~\bibnamefont {Pierleoni}},\
  and\ \bibinfo {author} {\bibfnamefont {D.~M.}\ \bibnamefont {Ceperley}},\
  }\bibfield  {title} {\bibinfo {title} {The properties of hydrogen and helium
  under extreme conditions},\ }\href
  {https://doi.org/10.1103/RevModPhys.84.1607} {\bibfield  {journal} {\bibinfo
  {journal} {Rev. Mod. Phys.}\ }\textbf {\bibinfo {volume} {84}},\ \bibinfo
  {pages} {1607} (\bibinfo {year} {2012})}\BibitemShut {NoStop}%
\bibitem [{\citenamefont {Bonitz}\ \emph {et~al.}(2024)\citenamefont {Bonitz},
  \citenamefont {Vorberger}, \citenamefont {Bethkenhagen}, \citenamefont
  {Böhme}, \citenamefont {Ceperley}, \citenamefont {Filinov}, \citenamefont
  {Gawne}, \citenamefont {Graziani}, \citenamefont {Gregori}, \citenamefont
  {Hamann}, \citenamefont {Hansen}, \citenamefont {Holzmann}, \citenamefont
  {Hu}, \citenamefont {Kählert}, \citenamefont {Karasiev}, \citenamefont
  {Kleinschmidt}, \citenamefont {Kordts}, \citenamefont {Makait}, \citenamefont
  {Militzer}, \citenamefont {Moldabekov}, \citenamefont {Pierleoni},
  \citenamefont {Preising}, \citenamefont {Ramakrishna}, \citenamefont
  {Redmer}, \citenamefont {Schwalbe}, \citenamefont {Svensson},\ and\
  \citenamefont {Dornheim}}]{2024_bonitz_hydrogen_simulation_review}%
  \BibitemOpen
  \bibfield  {author} {\bibinfo {author} {\bibfnamefont {M.}~\bibnamefont
  {Bonitz}}, \bibinfo {author} {\bibfnamefont {J.}~\bibnamefont {Vorberger}},
  \bibinfo {author} {\bibfnamefont {M.}~\bibnamefont {Bethkenhagen}}, \bibinfo
  {author} {\bibfnamefont {M.~P.}\ \bibnamefont {Böhme}}, \bibinfo {author}
  {\bibfnamefont {D.~M.}\ \bibnamefont {Ceperley}}, \bibinfo {author}
  {\bibfnamefont {A.}~\bibnamefont {Filinov}}, \bibinfo {author} {\bibfnamefont
  {T.}~\bibnamefont {Gawne}}, \bibinfo {author} {\bibfnamefont
  {F.}~\bibnamefont {Graziani}}, \bibinfo {author} {\bibfnamefont
  {G.}~\bibnamefont {Gregori}}, \bibinfo {author} {\bibfnamefont
  {P.}~\bibnamefont {Hamann}}, \bibinfo {author} {\bibfnamefont {S.~B.}\
  \bibnamefont {Hansen}}, \bibinfo {author} {\bibfnamefont {M.}~\bibnamefont
  {Holzmann}}, \bibinfo {author} {\bibfnamefont {S.~X.}\ \bibnamefont {Hu}},
  \bibinfo {author} {\bibfnamefont {H.}~\bibnamefont {Kählert}}, \bibinfo
  {author} {\bibfnamefont {V.~V.}\ \bibnamefont {Karasiev}}, \bibinfo {author}
  {\bibfnamefont {U.}~\bibnamefont {Kleinschmidt}}, \bibinfo {author}
  {\bibfnamefont {L.}~\bibnamefont {Kordts}}, \bibinfo {author} {\bibfnamefont
  {C.}~\bibnamefont {Makait}}, \bibinfo {author} {\bibfnamefont
  {B.}~\bibnamefont {Militzer}}, \bibinfo {author} {\bibfnamefont {Z.~A.}\
  \bibnamefont {Moldabekov}}, \bibinfo {author} {\bibfnamefont
  {C.}~\bibnamefont {Pierleoni}}, \bibinfo {author} {\bibfnamefont
  {M.}~\bibnamefont {Preising}}, \bibinfo {author} {\bibfnamefont
  {K.}~\bibnamefont {Ramakrishna}}, \bibinfo {author} {\bibfnamefont
  {R.}~\bibnamefont {Redmer}}, \bibinfo {author} {\bibfnamefont
  {S.}~\bibnamefont {Schwalbe}}, \bibinfo {author} {\bibfnamefont
  {P.}~\bibnamefont {Svensson}},\ and\ \bibinfo {author} {\bibfnamefont
  {T.}~\bibnamefont {Dornheim}},\ }\bibfield  {title} {\bibinfo {title} {Toward
  first principles-based simulations of dense hydrogen},\ }\href
  {https://doi.org/10.1063/5.0219405} {\bibfield  {journal} {\bibinfo
  {journal} {Physics of Plasmas}\ }\textbf {\bibinfo {volume} {31}},\ \bibinfo
  {pages} {110501} (\bibinfo {year} {2024})}\BibitemShut {NoStop}%
\bibitem [{\citenamefont {Ceperley}\ and\ \citenamefont
  {Alder}(1981)}]{1981_ceperley_metallic_hydrogen}%
  \BibitemOpen
  \bibfield  {author} {\bibinfo {author} {\bibfnamefont {D.}~\bibnamefont
  {Ceperley}}\ and\ \bibinfo {author} {\bibfnamefont {B.}~\bibnamefont
  {Alder}},\ }\bibfield  {title} {\bibinfo {title} {The calculation of the
  properties of metallic hydrogen using monte carlo},\ }\href
  {https://doi.org/https://doi.org/10.1016/0378-4363(81)90742-7} {\bibfield
  {journal} {\bibinfo  {journal} {Physica B+C}\ }\textbf {\bibinfo {volume}
  {108}},\ \bibinfo {pages} {875} (\bibinfo {year} {1981})}\BibitemShut
  {NoStop}%
\bibitem [{\citenamefont {Ceperley}\ and\ \citenamefont
  {Alder}(1987)}]{1987_ceperley_hydrogen_high_pressures}%
  \BibitemOpen
  \bibfield  {author} {\bibinfo {author} {\bibfnamefont {D.~M.}\ \bibnamefont
  {Ceperley}}\ and\ \bibinfo {author} {\bibfnamefont {B.~J.}\ \bibnamefont
  {Alder}},\ }\bibfield  {title} {\bibinfo {title} {{Ground state of solid
  hydrogen at high pressures}},\ }\href
  {https://doi.org/10.1103/PhysRevB.36.2092} {\bibfield  {journal} {\bibinfo
  {journal} {Physical Review B}\ }\textbf {\bibinfo {volume} {36}},\ \bibinfo
  {pages} {2092} (\bibinfo {year} {1987})}\BibitemShut {NoStop}%
\bibitem [{\citenamefont {Baroni}\ and\ \citenamefont
  {Moroni}(1999)}]{Baroni1999}%
  \BibitemOpen
  \bibfield  {author} {\bibinfo {author} {\bibfnamefont {S.}~\bibnamefont
  {Baroni}}\ and\ \bibinfo {author} {\bibfnamefont {S.}~\bibnamefont
  {Moroni}},\ }\bibfield  {title} {\bibinfo {title} {Reptation quantum monte
  carlo: A method for unbiased ground-state averages and imaginary-time
  correlations},\ }\href {https://doi.org/10.1103/PhysRevLett.82.4745}
  {\bibfield  {journal} {\bibinfo  {journal} {Phys. Rev. Lett.}\ }\textbf
  {\bibinfo {volume} {82}},\ \bibinfo {pages} {4745} (\bibinfo {year}
  {1999})}\BibitemShut {NoStop}%
\bibitem [{\citenamefont {Pierleoni}\ \emph {et~al.}(1994)\citenamefont
  {Pierleoni}, \citenamefont {Ceperley}, \citenamefont {Bernu},\ and\
  \citenamefont {Magro}}]{1994_pierleoni_hydrogen_plasma_pimc}%
  \BibitemOpen
  \bibfield  {author} {\bibinfo {author} {\bibfnamefont {C.}~\bibnamefont
  {Pierleoni}}, \bibinfo {author} {\bibfnamefont {D.}~\bibnamefont {Ceperley}},
  \bibinfo {author} {\bibfnamefont {B.}~\bibnamefont {Bernu}},\ and\ \bibinfo
  {author} {\bibfnamefont {W.}~\bibnamefont {Magro}},\ }\bibfield  {title}
  {\bibinfo {title} {{Equation of state of the hydrogen plasma by path integral
  Monte Carlo simulation}},\ }\href
  {papers://68dd8153-82a1-4a0e-9ae0-80ae65a14bf7/Paper/p158} {\bibfield
  {journal} {\bibinfo  {journal} {Physical Review Letters}\ }\textbf {\bibinfo
  {volume} {73}},\ \bibinfo {pages} {2145} (\bibinfo {year}
  {1994})}\BibitemShut {NoStop}%
\bibitem [{\citenamefont {Natoli}\ \emph {et~al.}(1993)\citenamefont {Natoli},
  \citenamefont {Martin},\ and\ \citenamefont
  {Ceperley}}]{1993_natoli_solid_hydrogen}%
  \BibitemOpen
  \bibfield  {author} {\bibinfo {author} {\bibfnamefont {V.}~\bibnamefont
  {Natoli}}, \bibinfo {author} {\bibfnamefont {R.~M.}\ \bibnamefont {Martin}},\
  and\ \bibinfo {author} {\bibfnamefont {D.~M.}\ \bibnamefont {Ceperley}},\
  }\bibfield  {title} {\bibinfo {title} {Crystal structure of atomic
  hydrogen},\ }\href {https://doi.org/10.1103/PhysRevLett.70.1952} {\bibfield
  {journal} {\bibinfo  {journal} {Phys. Rev. Lett.}\ }\textbf {\bibinfo
  {volume} {70}},\ \bibinfo {pages} {1952} (\bibinfo {year}
  {1993})}\BibitemShut {NoStop}%
\bibitem [{\citenamefont {Natoli}\ \emph {et~al.}(1995)\citenamefont {Natoli},
  \citenamefont {Martin},\ and\ \citenamefont {Ceperley}}]{Natoli95}%
  \BibitemOpen
  \bibfield  {author} {\bibinfo {author} {\bibfnamefont {V.}~\bibnamefont
  {Natoli}}, \bibinfo {author} {\bibfnamefont {R.~M.}\ \bibnamefont {Martin}},\
  and\ \bibinfo {author} {\bibfnamefont {D.}~\bibnamefont {Ceperley}},\
  }\bibfield  {title} {\bibinfo {title} {Crystal structure of molecular
  hydrogen at high pressure},\ }\href
  {https://doi.org/10.1103/PhysRevLett.74.1601} {\bibfield  {journal} {\bibinfo
   {journal} {Phys. Rev. Lett.}\ }\textbf {\bibinfo {volume} {74}},\ \bibinfo
  {pages} {1601} (\bibinfo {year} {1995})}\BibitemShut {NoStop}%
\bibitem [{\citenamefont {Holzmann}\ \emph {et~al.}(2005)\citenamefont
  {Holzmann}, \citenamefont {Pierleoni},\ and\ \citenamefont
  {Ceperley}}]{2005_holzmann_hydrogen}%
  \BibitemOpen
  \bibfield  {author} {\bibinfo {author} {\bibfnamefont {M.}~\bibnamefont
  {Holzmann}}, \bibinfo {author} {\bibfnamefont {C.}~\bibnamefont
  {Pierleoni}},\ and\ \bibinfo {author} {\bibfnamefont {D.~M.}\ \bibnamefont
  {Ceperley}},\ }\bibfield  {title} {\bibinfo {title} {Coupled electron–ion
  monte carlo calculations of atomic hydrogen},\ }\href
  {https://doi.org/https://doi.org/10.1016/j.cpc.2005.03.093} {\bibfield
  {journal} {\bibinfo  {journal} {Computer Physics Communications}\ }\textbf
  {\bibinfo {volume} {169}},\ \bibinfo {pages} {421} (\bibinfo {year}
  {2005})},\ \bibinfo {note} {proceedings of the Europhysics Conference on
  Computational Physics 2004}\BibitemShut {NoStop}%
\bibitem [{\citenamefont {Pierleoni}\ \emph {et~al.}(2004)\citenamefont
  {Pierleoni}, \citenamefont {Ceperley},\ and\ \citenamefont
  {Holzmann}}]{2004_pierleoni_ceimc_hydrogen}%
  \BibitemOpen
  \bibfield  {author} {\bibinfo {author} {\bibfnamefont {C.}~\bibnamefont
  {Pierleoni}}, \bibinfo {author} {\bibfnamefont {D.}~\bibnamefont
  {Ceperley}},\ and\ \bibinfo {author} {\bibfnamefont {M.}~\bibnamefont
  {Holzmann}},\ }\bibfield  {title} {\bibinfo {title} {{Coupled Electron-Ion
  Monte Carlo calculations of dense metallic hydrogen}},\ }\href
  {papers://68dd8153-82a1-4a0e-9ae0-80ae65a14bf7/Paper/p192} {\bibfield
  {journal} {\bibinfo  {journal} {Physical Review Letters}\ }\textbf {\bibinfo
  {volume} {93}},\ \bibinfo {pages} {146402} (\bibinfo {year}
  {2004})}\BibitemShut {NoStop}%
\bibitem [{\citenamefont {Attaccalite}\ and\ \citenamefont
  {Sorella}(2008)}]{2008_attaccalite_md_hydrogen}%
  \BibitemOpen
  \bibfield  {author} {\bibinfo {author} {\bibfnamefont {C.}~\bibnamefont
  {Attaccalite}}\ and\ \bibinfo {author} {\bibfnamefont {S.}~\bibnamefont
  {Sorella}},\ }\bibfield  {title} {\bibinfo {title} {Stable liquid hydrogen at
  high pressure by a novel ab initio molecular-dynamics calculation},\ }\href
  {https://doi.org/10.1103/PhysRevLett.100.114501} {\bibfield  {journal}
  {\bibinfo  {journal} {Phys. Rev. Lett.}\ }\textbf {\bibinfo {volume} {100}},\
  \bibinfo {pages} {114501} (\bibinfo {year} {2008})}\BibitemShut {NoStop}%
\bibitem [{\citenamefont {Wang}\ \emph {et~al.}(1990)\citenamefont {Wang},
  \citenamefont {Zhu}, \citenamefont {Louie},\ and\ \citenamefont
  {Fahy}}]{Wang90}%
  \BibitemOpen
  \bibfield  {author} {\bibinfo {author} {\bibfnamefont {X.~W.}\ \bibnamefont
  {Wang}}, \bibinfo {author} {\bibfnamefont {J.}~\bibnamefont {Zhu}}, \bibinfo
  {author} {\bibfnamefont {S.~G.}\ \bibnamefont {Louie}},\ and\ \bibinfo
  {author} {\bibfnamefont {S.}~\bibnamefont {Fahy}},\ }\bibfield  {title}
  {\bibinfo {title} {Magnetic structure and equation of state of bcc solid
  hydrogen: A variational quantum monte carlo study},\ }\href
  {https://doi.org/10.1103/PhysRevLett.65.2414} {\bibfield  {journal} {\bibinfo
   {journal} {Phys. Rev. Lett.}\ }\textbf {\bibinfo {volume} {65}},\ \bibinfo
  {pages} {2414} (\bibinfo {year} {1990})}\BibitemShut {NoStop}%
\bibitem [{\citenamefont {Pierleoni}\ \emph
  {et~al.}(2008{\natexlab{a}})\citenamefont {Pierleoni}, \citenamefont
  {Delaney}, \citenamefont {Morales}, \citenamefont {Ceperley},\ and\
  \citenamefont {Holzmann}}]{Pierleoni08}%
  \BibitemOpen
  \bibfield  {author} {\bibinfo {author} {\bibfnamefont {C.}~\bibnamefont
  {Pierleoni}}, \bibinfo {author} {\bibfnamefont {K.~T.}\ \bibnamefont
  {Delaney}}, \bibinfo {author} {\bibfnamefont {M.~A.}\ \bibnamefont
  {Morales}}, \bibinfo {author} {\bibfnamefont {D.~M.}\ \bibnamefont
  {Ceperley}},\ and\ \bibinfo {author} {\bibfnamefont {M.}~\bibnamefont
  {Holzmann}},\ }\bibfield  {title} {\bibinfo {title} {Trial wave functions for
  high-pressure metallic hydrogen},\ }\href
  {https://doi.org/https://doi.org/10.1016/j.cpc.2008.01.041} {\bibfield
  {journal} {\bibinfo  {journal} {Computer Physics Communications}\ }\textbf
  {\bibinfo {volume} {179}},\ \bibinfo {pages} {89} (\bibinfo {year}
  {2008}{\natexlab{a}})},\ \bibinfo {note} {special issue based on the
  Conference on Computational Physics 2007}\BibitemShut {NoStop}%
\bibitem [{\citenamefont {Holzmann}\ \emph {et~al.}(2003)\citenamefont
  {Holzmann}, \citenamefont {Ceperley}, \citenamefont {Pierleoni},\ and\
  \citenamefont {Esler}}]{2003_holzmann_backflow_electron_gas_and_hydrogen}%
  \BibitemOpen
  \bibfield  {author} {\bibinfo {author} {\bibfnamefont {M.}~\bibnamefont
  {Holzmann}}, \bibinfo {author} {\bibfnamefont {D.~M.}\ \bibnamefont
  {Ceperley}}, \bibinfo {author} {\bibfnamefont {C.}~\bibnamefont
  {Pierleoni}},\ and\ \bibinfo {author} {\bibfnamefont {K.}~\bibnamefont
  {Esler}},\ }\bibfield  {title} {\bibinfo {title} {Backflow correlations for
  the electron gas and metallic hydrogen},\ }\href
  {https://doi.org/10.1103/PhysRevE.68.046707} {\bibfield  {journal} {\bibinfo
  {journal} {Phys. Rev. E}\ }\textbf {\bibinfo {volume} {68}},\ \bibinfo
  {pages} {046707} (\bibinfo {year} {2003})}\BibitemShut {NoStop}%
\bibitem [{\citenamefont {Pierleoni}\ and\ \citenamefont
  {Ceperley}(2006)}]{Pierleoni2006}%
  \BibitemOpen
  \bibfield  {author} {\bibinfo {author} {\bibfnamefont {C.}~\bibnamefont
  {Pierleoni}}\ and\ \bibinfo {author} {\bibfnamefont {D.}~\bibnamefont
  {Ceperley}},\ }\bibinfo {title} {The coupled electron-ion monte carlo
  method},\ in\ \href {https://doi.org/10.1007/3-540-35273-2_18} {\emph
  {\bibinfo {booktitle} {Computer Simulations in Condensed Matter Systems: From
  Materials to Chemical Biology Volume 1}}},\ \bibinfo {editor} {edited by\
  \bibinfo {editor} {\bibfnamefont {M.}~\bibnamefont {Ferrario}}, \bibinfo
  {editor} {\bibfnamefont {G.}~\bibnamefont {Ciccotti}},\ and\ \bibinfo
  {editor} {\bibfnamefont {K.}~\bibnamefont {Binder}}}\ (\bibinfo  {publisher}
  {Springer Berlin Heidelberg},\ \bibinfo {address} {Berlin, Heidelberg},\
  \bibinfo {year} {2006})\ pp.\ \bibinfo {pages} {641--683}\BibitemShut
  {NoStop}%
\bibitem [{\citenamefont {Holzmann}(2024)}]{Holzmann24}%
  \BibitemOpen
  \bibfield  {author} {\bibinfo {author} {\bibfnamefont {M.}~\bibnamefont
  {Holzmann}},\ }\bibinfo {title} {High-pressure phases of hydrogen},\ in\
  \href@noop {} {\emph {\bibinfo {booktitle} {Correlations and Phase
  Transitions}}},\ \bibinfo {editor} {edited by\ \bibinfo {editor}
  {\bibfnamefont {E.}~\bibnamefont {Pavarini}}\ and\ \bibinfo {editor}
  {\bibfnamefont {E.}~\bibnamefont {Koch}}}\ (\bibinfo  {publisher} {Verlag des
  Forschungszentrums Jülich},\ \bibinfo {address} {Jülich},\ \bibinfo {year}
  {2024})\BibitemShut {NoStop}%
\bibitem [{\citenamefont {Xie}\ \emph {et~al.}(2023)\citenamefont {Xie},
  \citenamefont {Li}, \citenamefont {Wang}, \citenamefont {Zhang},\ and\
  \citenamefont {Wang}}]{Xie23}%
  \BibitemOpen
  \bibfield  {author} {\bibinfo {author} {\bibfnamefont {H.}~\bibnamefont
  {Xie}}, \bibinfo {author} {\bibfnamefont {Z.-H.}\ \bibnamefont {Li}},
  \bibinfo {author} {\bibfnamefont {H.}~\bibnamefont {Wang}}, \bibinfo {author}
  {\bibfnamefont {L.}~\bibnamefont {Zhang}},\ and\ \bibinfo {author}
  {\bibfnamefont {L.}~\bibnamefont {Wang}},\ }\bibfield  {title} {\bibinfo
  {title} {Deep variational free energy approach to dense hydrogen},\ }\href
  {https://doi.org/10.1103/PhysRevLett.131.126501} {\bibfield  {journal}
  {\bibinfo  {journal} {Phys. Rev. Lett.}\ }\textbf {\bibinfo {volume} {131}},\
  \bibinfo {pages} {126501} (\bibinfo {year} {2023})}\BibitemShut {NoStop}%
\bibitem [{\citenamefont {Dong}\ \emph {et~al.}(2025)\citenamefont {Dong},
  \citenamefont {Xie}, \citenamefont {Chen}, \citenamefont {Liang},
  \citenamefont {Zhang}, \citenamefont {Wang},\ and\ \citenamefont
  {Wang}}]{Dong25}%
  \BibitemOpen
  \bibfield  {author} {\bibinfo {author} {\bibfnamefont {X.}~\bibnamefont
  {Dong}}, \bibinfo {author} {\bibfnamefont {H.}~\bibnamefont {Xie}}, \bibinfo
  {author} {\bibfnamefont {Y.}~\bibnamefont {Chen}}, \bibinfo {author}
  {\bibfnamefont {W.}~\bibnamefont {Liang}}, \bibinfo {author} {\bibfnamefont
  {L.}~\bibnamefont {Zhang}}, \bibinfo {author} {\bibfnamefont
  {L.}~\bibnamefont {Wang}},\ and\ \bibinfo {author} {\bibfnamefont
  {H.}~\bibnamefont {Wang}},\ }\bibfield  {title} {\bibinfo {title} {Deep
  variational free energy prediction of dense hydrogen solid at 1200 {K}},\
  }\href {https://doi.org/10.1103/rbsg-r7hx} {\bibfield  {journal} {\bibinfo
  {journal} {Phys. Rev. B}\ }\textbf {\bibinfo {volume} {111}},\ \bibinfo
  {pages} {214118} (\bibinfo {year} {2025})}\BibitemShut {NoStop}%
\bibitem [{\citenamefont {Ruggeri}\ \emph {et~al.}(2020)\citenamefont
  {Ruggeri}, \citenamefont {Holzmann}, \citenamefont {Ceperley},\ and\
  \citenamefont {Pierleoni}}]{Ruggeri2020}%
  \BibitemOpen
  \bibfield  {author} {\bibinfo {author} {\bibfnamefont {M.}~\bibnamefont
  {Ruggeri}}, \bibinfo {author} {\bibfnamefont {M.}~\bibnamefont {Holzmann}},
  \bibinfo {author} {\bibfnamefont {D.~M.}\ \bibnamefont {Ceperley}},\ and\
  \bibinfo {author} {\bibfnamefont {C.}~\bibnamefont {Pierleoni}},\ }\bibfield
  {title} {\bibinfo {title} {Quantum monte carlo determination of the principal
  hugoniot of deuterium},\ }\href {https://doi.org/10.1103/PhysRevB.102.144108}
  {\bibfield  {journal} {\bibinfo  {journal} {Phys. Rev. B}\ }\textbf {\bibinfo
  {volume} {102}},\ \bibinfo {pages} {144108} (\bibinfo {year}
  {2020})}\BibitemShut {NoStop}%
\bibitem [{\citenamefont {Tirelli}\ \emph {et~al.}(2022)\citenamefont
  {Tirelli}, \citenamefont {Tenti}, \citenamefont {Nakano},\ and\ \citenamefont
  {Sorella}}]{Tirelli2022}%
  \BibitemOpen
  \bibfield  {author} {\bibinfo {author} {\bibfnamefont {A.}~\bibnamefont
  {Tirelli}}, \bibinfo {author} {\bibfnamefont {G.}~\bibnamefont {Tenti}},
  \bibinfo {author} {\bibfnamefont {K.}~\bibnamefont {Nakano}},\ and\ \bibinfo
  {author} {\bibfnamefont {S.}~\bibnamefont {Sorella}},\ }\bibfield  {title}
  {\bibinfo {title} {High-pressure hydrogen by machine learning and quantum
  monte carlo},\ }\href {https://doi.org/10.1103/PhysRevB.106.L041105}
  {\bibfield  {journal} {\bibinfo  {journal} {Phys. Rev. B}\ }\textbf {\bibinfo
  {volume} {106}},\ \bibinfo {pages} {L041105} (\bibinfo {year}
  {2022})}\BibitemShut {NoStop}%
\bibitem [{\citenamefont {Niu}\ \emph {et~al.}(2023)\citenamefont {Niu},
  \citenamefont {Yang}, \citenamefont {Jensen}, \citenamefont {Holzmann},
  \citenamefont {Pierleoni},\ and\ \citenamefont
  {Ceperley}}]{2023_niu_qmc_database_hydrogen}%
  \BibitemOpen
  \bibfield  {author} {\bibinfo {author} {\bibfnamefont {H.}~\bibnamefont
  {Niu}}, \bibinfo {author} {\bibfnamefont {Y.}~\bibnamefont {Yang}}, \bibinfo
  {author} {\bibfnamefont {S.}~\bibnamefont {Jensen}}, \bibinfo {author}
  {\bibfnamefont {M.}~\bibnamefont {Holzmann}}, \bibinfo {author}
  {\bibfnamefont {C.}~\bibnamefont {Pierleoni}},\ and\ \bibinfo {author}
  {\bibfnamefont {D.~M.}\ \bibnamefont {Ceperley}},\ }\bibfield  {title}
  {\bibinfo {title} {Stable solid molecular hydrogen above 900 {K} from a
  machine-learned potential trained with diffusion quantum monte carlo},\
  }\href {https://doi.org/10.1103/PhysRevLett.130.076102} {\bibfield  {journal}
  {\bibinfo  {journal} {Phys. Rev. Lett.}\ }\textbf {\bibinfo {volume} {130}},\
  \bibinfo {pages} {076102} (\bibinfo {year} {2023})}\BibitemShut {NoStop}%
\bibitem [{\citenamefont {Goswami}\ \emph {et~al.}(2025)\citenamefont
  {Goswami}, \citenamefont {Jensen}, \citenamefont {Yang}, \citenamefont
  {Holzmann}, \citenamefont {Pierleoni},\ and\ \citenamefont
  {Ceperley}}]{goswami2024high}%
  \BibitemOpen
  \bibfield  {author} {\bibinfo {author} {\bibfnamefont {S.}~\bibnamefont
  {Goswami}}, \bibinfo {author} {\bibfnamefont {S.}~\bibnamefont {Jensen}},
  \bibinfo {author} {\bibfnamefont {Y.}~\bibnamefont {Yang}}, \bibinfo {author}
  {\bibfnamefont {M.}~\bibnamefont {Holzmann}}, \bibinfo {author}
  {\bibfnamefont {C.}~\bibnamefont {Pierleoni}},\ and\ \bibinfo {author}
  {\bibfnamefont {D.~M.}\ \bibnamefont {Ceperley}},\ }\bibfield  {title}
  {\bibinfo {title} {High temperature melting of dense molecular hydrogen from
  machine-learning interatomic potentials trained on quantum monte carlo},\
  }\href {https://doi.org/https://doi.org/10.1063/5.0250686} {\bibfield
  {journal} {\bibinfo  {journal} {The Journal of Chemical Physics}\ }\textbf
  {\bibinfo {volume} {162}},\ \bibinfo {pages} {054118} (\bibinfo {year}
  {2025})}\BibitemShut {NoStop}%
\bibitem [{\citenamefont {Ly}\ and\ \citenamefont {Ceperley}(2025)}]{Ly2025}%
  \BibitemOpen
  \bibfield  {author} {\bibinfo {author} {\bibfnamefont {K.~K.}\ \bibnamefont
  {Ly}}\ and\ \bibinfo {author} {\bibfnamefont {D.~M.}\ \bibnamefont
  {Ceperley}},\ }\bibfield  {title} {\bibinfo {title} {Melting curves of atomic
  hydrogen and deuterium calculated using path-integral monte carlo},\ }\href
  {https://doi.org/10.1103/PhysRevB.111.104102} {\bibfield  {journal} {\bibinfo
   {journal} {Phys. Rev. B}\ }\textbf {\bibinfo {volume} {111}},\ \bibinfo
  {pages} {104102} (\bibinfo {year} {2025})}\BibitemShut {NoStop}%
\bibitem [{\citenamefont {Holzmann}\ and\ \citenamefont
  {Moroni}(2019)}]{2019_holzmann_backflow}%
  \BibitemOpen
  \bibfield  {author} {\bibinfo {author} {\bibfnamefont {M.}~\bibnamefont
  {Holzmann}}\ and\ \bibinfo {author} {\bibfnamefont {S.}~\bibnamefont
  {Moroni}},\ }\bibfield  {title} {\bibinfo {title} {Orbital-dependent backflow
  wave functions for real-space quantum monte carlo},\ }\href
  {https://doi.org/10.1103/PhysRevB.99.085121} {\bibfield  {journal} {\bibinfo
  {journal} {Phys. Rev. B}\ }\textbf {\bibinfo {volume} {99}},\ \bibinfo
  {pages} {085121} (\bibinfo {year} {2019})}\BibitemShut {NoStop}%
\bibitem [{\citenamefont {Pfau}\ \emph {et~al.}(2020)\citenamefont {Pfau},
  \citenamefont {Spencer}, \citenamefont {Matthews},\ and\ \citenamefont
  {Foulkes}}]{2020_pfau_deep_net}%
  \BibitemOpen
  \bibfield  {author} {\bibinfo {author} {\bibfnamefont {D.}~\bibnamefont
  {Pfau}}, \bibinfo {author} {\bibfnamefont {J.~S.}\ \bibnamefont {Spencer}},
  \bibinfo {author} {\bibfnamefont {A.~G.}\ \bibnamefont {Matthews}},\ and\
  \bibinfo {author} {\bibfnamefont {W.~M.~C.}\ \bibnamefont {Foulkes}},\
  }\bibfield  {title} {\bibinfo {title} {Ab initio solution of the
  many-electron {S}chr{\"o}dinger equation with deep neural networks},\
  }\href@noop {} {\bibfield  {journal} {\bibinfo  {journal} {Physical review
  research}\ }\textbf {\bibinfo {volume} {2}},\ \bibinfo {pages} {033429}
  (\bibinfo {year} {2020})}\BibitemShut {NoStop}%
\bibitem [{\citenamefont {Hermann}\ \emph {et~al.}(2020)\citenamefont
  {Hermann}, \citenamefont {Sch{\"a}tzle},\ and\ \citenamefont
  {No{\'e}}}]{2020_hermann_neural_net_electrons}%
  \BibitemOpen
  \bibfield  {author} {\bibinfo {author} {\bibfnamefont {J.}~\bibnamefont
  {Hermann}}, \bibinfo {author} {\bibfnamefont {Z.}~\bibnamefont
  {Sch{\"a}tzle}},\ and\ \bibinfo {author} {\bibfnamefont {F.}~\bibnamefont
  {No{\'e}}},\ }\bibfield  {title} {\bibinfo {title} {Deep-neural-network
  solution of the electronic {S}chr{\"o}dinger equation},\ }\href@noop {}
  {\bibfield  {journal} {\bibinfo  {journal} {Nature Chemistry}\ }\textbf
  {\bibinfo {volume} {12}},\ \bibinfo {pages} {891} (\bibinfo {year}
  {2020})}\BibitemShut {NoStop}%
\bibitem [{\citenamefont {Gao}\ and\ \citenamefont
  {G{\"u}nnemann}(2023)}]{2023_gao_generalizing_neural_wavefunctions}%
  \BibitemOpen
  \bibfield  {author} {\bibinfo {author} {\bibfnamefont {N.}~\bibnamefont
  {Gao}}\ and\ \bibinfo {author} {\bibfnamefont {S.}~\bibnamefont
  {G{\"u}nnemann}},\ }\bibfield  {title} {\bibinfo {title} {Generalizing neural
  wave functions},\ }in\ \href@noop {} {\emph {\bibinfo {booktitle}
  {Proceedings of the 40th International Conference on Machine Learning}}},\
  \bibinfo {series} {Proceedings of Machine Learning Research}, Vol.\ \bibinfo
  {volume} {202},\ \bibinfo {editor} {edited by\ \bibinfo {editor}
  {\bibfnamefont {A.}~\bibnamefont {Krause}}, \bibinfo {editor} {\bibfnamefont
  {E.}~\bibnamefont {Brunskill}}, \bibinfo {editor} {\bibfnamefont
  {K.}~\bibnamefont {Cho}}, \bibinfo {editor} {\bibfnamefont {B.}~\bibnamefont
  {Engelhardt}}, \bibinfo {editor} {\bibfnamefont {S.}~\bibnamefont {Sabato}},\
  and\ \bibinfo {editor} {\bibfnamefont {J.}~\bibnamefont {Scarlett}}}\
  (\bibinfo  {publisher} {PMLR},\ \bibinfo {year} {2023})\ pp.\ \bibinfo
  {pages} {10708--10726}\BibitemShut {NoStop}%
\bibitem [{\citenamefont {Gao}\ and\ \citenamefont {Günnemann}(2024)}]{Gao24}%
  \BibitemOpen
  \bibfield  {author} {\bibinfo {author} {\bibfnamefont {N.}~\bibnamefont
  {Gao}}\ and\ \bibinfo {author} {\bibfnamefont {S.}~\bibnamefont
  {Günnemann}},\ }\href {https://arxiv.org/abs/2405.14762} {\bibinfo {title}
  {Neural pfaffians: Solving many many-electron {S}chr\"odinger equations}}
  (\bibinfo {year} {2024}),\ \Eprint {https://arxiv.org/abs/2405.14762}
  {arXiv:2405.14762 [cs.LG]} \BibitemShut {NoStop}%
\bibitem [{\citenamefont {Scherbela}\ \emph {et~al.}(2024)\citenamefont
  {Scherbela}, \citenamefont {Gerard},\ and\ \citenamefont
  {Grohs}}]{2024_scherbela_transferable_fermionic_neural_wf}%
  \BibitemOpen
  \bibfield  {author} {\bibinfo {author} {\bibfnamefont {M.}~\bibnamefont
  {Scherbela}}, \bibinfo {author} {\bibfnamefont {L.}~\bibnamefont {Gerard}},\
  and\ \bibinfo {author} {\bibfnamefont {P.}~\bibnamefont {Grohs}},\ }\bibfield
   {title} {\bibinfo {title} {Towards a transferable fermionic neural
  wavefunction for molecules},\ }\href
  {https://doi.org/10.1038/s41467-023-44216-9} {\bibfield  {journal} {\bibinfo
  {journal} {Nature Communications}\ }\textbf {\bibinfo {volume} {15}},\
  \bibinfo {pages} {120} (\bibinfo {year} {2024})}\BibitemShut {NoStop}%
\bibitem [{\citenamefont {Zhang}\ \emph {et~al.}(2025)\citenamefont {Zhang},
  \citenamefont {Jiang},\ and\ \citenamefont {Guo}}]{Zhang25}%
  \BibitemOpen
  \bibfield  {author} {\bibinfo {author} {\bibfnamefont {Y.}~\bibnamefont
  {Zhang}}, \bibinfo {author} {\bibfnamefont {B.}~\bibnamefont {Jiang}},\ and\
  \bibinfo {author} {\bibfnamefont {H.}~\bibnamefont {Guo}},\ }\bibfield
  {title} {\bibinfo {title} {Schr{\"o}dingernet: A universal neural network
  solver for the {S}chr{\"o}dinger equation},\ }\href
  {https://doi.org/10.1021/acs.jctc.4c01287} {\bibfield  {journal} {\bibinfo
  {journal} {Journal of Chemical Theory and Computation}\ }\textbf {\bibinfo
  {volume} {21}},\ \bibinfo {pages} {670} (\bibinfo {year} {2025})},\ \bibinfo
  {note} {pMID: 39772624},\ \Eprint
  {https://arxiv.org/abs/https://doi.org/10.1021/acs.jctc.4c01287}
  {https://doi.org/10.1021/acs.jctc.4c01287} \BibitemShut {NoStop}%
\bibitem [{\citenamefont {Hermann}\ \emph {et~al.}(2023)\citenamefont
  {Hermann}, \citenamefont {Spencer}, \citenamefont {Choo}, \citenamefont
  {Mezzacapo}, \citenamefont {Foulkes}, \citenamefont {Pfau}, \citenamefont
  {Carleo},\ and\ \citenamefont {Noé}}]{hermann_ab_2023}%
  \BibitemOpen
  \bibfield  {author} {\bibinfo {author} {\bibfnamefont {J.}~\bibnamefont
  {Hermann}}, \bibinfo {author} {\bibfnamefont {J.}~\bibnamefont {Spencer}},
  \bibinfo {author} {\bibfnamefont {K.}~\bibnamefont {Choo}}, \bibinfo {author}
  {\bibfnamefont {A.}~\bibnamefont {Mezzacapo}}, \bibinfo {author}
  {\bibfnamefont {W.~M.~C.}\ \bibnamefont {Foulkes}}, \bibinfo {author}
  {\bibfnamefont {D.}~\bibnamefont {Pfau}}, \bibinfo {author} {\bibfnamefont
  {G.}~\bibnamefont {Carleo}},\ and\ \bibinfo {author} {\bibfnamefont
  {F.}~\bibnamefont {Noé}},\ }\bibfield  {title} {\bibinfo {title} {Ab initio
  quantum chemistry with neural-network wavefunctions},\ }\href
  {https://doi.org/10.1038/s41570-023-00516-8} {\bibfield  {journal} {\bibinfo
  {journal} {Nature Reviews Chemistry}\ }\textbf {\bibinfo {volume} {7}},\
  \bibinfo {pages} {692} (\bibinfo {year} {2023})}\BibitemShut {NoStop}%
\bibitem [{\citenamefont {Scherbela}\ \emph {et~al.}(2025)\citenamefont
  {Scherbela}, \citenamefont {Gao}, \citenamefont {Grohs},\ and\ \citenamefont
  {Günnemann}}]{scherbela_accurate_2025}%
  \BibitemOpen
  \bibfield  {author} {\bibinfo {author} {\bibfnamefont {M.}~\bibnamefont
  {Scherbela}}, \bibinfo {author} {\bibfnamefont {N.}~\bibnamefont {Gao}},
  \bibinfo {author} {\bibfnamefont {P.}~\bibnamefont {Grohs}},\ and\ \bibinfo
  {author} {\bibfnamefont {S.}~\bibnamefont {Günnemann}},\ }\href
  {https://doi.org/10.48550/arXiv.2504.06087} {\bibinfo {title} {Accurate
  {Ab}-initio {Neural}-network {Solutions} to {Large}-{Scale} {Electronic}
  {Structure} {Problems}}} (\bibinfo {year} {2025}),\ \bibinfo {note}
  {arXiv:2504.06087 [physics]}\BibitemShut {NoStop}%
\bibitem [{\citenamefont {Holzmann}\ and\ \citenamefont
  {Moroni}(2020)}]{Holzmann20}%
  \BibitemOpen
  \bibfield  {author} {\bibinfo {author} {\bibfnamefont {M.}~\bibnamefont
  {Holzmann}}\ and\ \bibinfo {author} {\bibfnamefont {S.}~\bibnamefont
  {Moroni}},\ }\bibfield  {title} {\bibinfo {title} {Itinerant-electron
  magnetism: The importance of many-body correlations},\ }\href
  {https://doi.org/10.1103/PhysRevLett.124.206404} {\bibfield  {journal}
  {\bibinfo  {journal} {Phys. Rev. Lett.}\ }\textbf {\bibinfo {volume} {124}},\
  \bibinfo {pages} {206404} (\bibinfo {year} {2020})}\BibitemShut {NoStop}%
\bibitem [{\citenamefont {Wilson}\ \emph {et~al.}(2023)\citenamefont {Wilson},
  \citenamefont {Moroni}, \citenamefont {Holzmann}, \citenamefont {Gao},
  \citenamefont {Wudarski}, \citenamefont {Vegge},\ and\ \citenamefont
  {Bhowmik}}]{Wilson23}%
  \BibitemOpen
  \bibfield  {author} {\bibinfo {author} {\bibfnamefont {M.}~\bibnamefont
  {Wilson}}, \bibinfo {author} {\bibfnamefont {S.}~\bibnamefont {Moroni}},
  \bibinfo {author} {\bibfnamefont {M.}~\bibnamefont {Holzmann}}, \bibinfo
  {author} {\bibfnamefont {N.}~\bibnamefont {Gao}}, \bibinfo {author}
  {\bibfnamefont {F.}~\bibnamefont {Wudarski}}, \bibinfo {author}
  {\bibfnamefont {T.}~\bibnamefont {Vegge}},\ and\ \bibinfo {author}
  {\bibfnamefont {A.}~\bibnamefont {Bhowmik}},\ }\bibfield  {title} {\bibinfo
  {title} {Neural network ansatz for periodic wave functions and the
  homogeneous electron gas},\ }\href
  {https://doi.org/10.1103/PhysRevB.107.235139} {\bibfield  {journal} {\bibinfo
   {journal} {Phys. Rev. B}\ }\textbf {\bibinfo {volume} {107}},\ \bibinfo
  {pages} {235139} (\bibinfo {year} {2023})}\BibitemShut {NoStop}%
\bibitem [{\citenamefont {Li}\ \emph {et~al.}(2022)\citenamefont {Li},
  \citenamefont {Li},\ and\ \citenamefont {Chen}}]{LiLi}%
  \BibitemOpen
  \bibfield  {author} {\bibinfo {author} {\bibfnamefont {X.}~\bibnamefont
  {Li}}, \bibinfo {author} {\bibfnamefont {Z.}~\bibnamefont {Li}},\ and\
  \bibinfo {author} {\bibfnamefont {J.}~\bibnamefont {Chen}},\ }\bibfield
  {title} {\bibinfo {title} {Ab initio calculation of real solids via neural
  network ansatz},\ }\href {https://doi.org/10.1038/s41467-022-35627-1}
  {\bibfield  {journal} {\bibinfo  {journal} {Nature Communications}\ }\textbf
  {\bibinfo {volume} {13}},\ \bibinfo {pages} {7895} (\bibinfo {year}
  {2022})}\BibitemShut {NoStop}%
\bibitem [{\citenamefont {Cassella}\ \emph {et~al.}(2023)\citenamefont
  {Cassella}, \citenamefont {Sutterud}, \citenamefont {Azadi}, \citenamefont
  {Drummond}, \citenamefont {Pfau}, \citenamefont {Spencer},\ and\
  \citenamefont {Foulkes}}]{Cassella23}%
  \BibitemOpen
  \bibfield  {author} {\bibinfo {author} {\bibfnamefont {G.}~\bibnamefont
  {Cassella}}, \bibinfo {author} {\bibfnamefont {H.}~\bibnamefont {Sutterud}},
  \bibinfo {author} {\bibfnamefont {S.}~\bibnamefont {Azadi}}, \bibinfo
  {author} {\bibfnamefont {N.~D.}\ \bibnamefont {Drummond}}, \bibinfo {author}
  {\bibfnamefont {D.}~\bibnamefont {Pfau}}, \bibinfo {author} {\bibfnamefont
  {J.~S.}\ \bibnamefont {Spencer}},\ and\ \bibinfo {author} {\bibfnamefont
  {W.~M.~C.}\ \bibnamefont {Foulkes}},\ }\bibfield  {title} {\bibinfo {title}
  {Discovering quantum phase transitions with fermionic neural networks},\
  }\href {https://doi.org/10.1103/PhysRevLett.130.036401} {\bibfield  {journal}
  {\bibinfo  {journal} {Phys. Rev. Lett.}\ }\textbf {\bibinfo {volume} {130}},\
  \bibinfo {pages} {036401} (\bibinfo {year} {2023})}\BibitemShut {NoStop}%
\bibitem [{\citenamefont {Smith}\ \emph {et~al.}(2024)\citenamefont {Smith},
  \citenamefont {Chen}, \citenamefont {Levy}, \citenamefont {Yang},
  \citenamefont {Morales},\ and\ \citenamefont {Zhang}}]{Smith24}%
  \BibitemOpen
  \bibfield  {author} {\bibinfo {author} {\bibfnamefont {C.}~\bibnamefont
  {Smith}}, \bibinfo {author} {\bibfnamefont {Y.}~\bibnamefont {Chen}},
  \bibinfo {author} {\bibfnamefont {R.}~\bibnamefont {Levy}}, \bibinfo {author}
  {\bibfnamefont {Y.}~\bibnamefont {Yang}}, \bibinfo {author} {\bibfnamefont
  {M.~A.}\ \bibnamefont {Morales}},\ and\ \bibinfo {author} {\bibfnamefont
  {S.}~\bibnamefont {Zhang}},\ }\bibfield  {title} {\bibinfo {title} {Unified
  variational approach description of ground-state phases of the
  two-dimensional electron gas},\ }\href
  {https://doi.org/10.1103/PhysRevLett.133.266504} {\bibfield  {journal}
  {\bibinfo  {journal} {Phys. Rev. Lett.}\ }\textbf {\bibinfo {volume} {133}},\
  \bibinfo {pages} {266504} (\bibinfo {year} {2024})}\BibitemShut {NoStop}%
\bibitem [{\citenamefont {Luo}\ \emph {et~al.}(2024)\citenamefont {Luo},
  \citenamefont {Dai},\ and\ \citenamefont {Fu}}]{luo_simulating_2024}%
  \BibitemOpen
  \bibfield  {author} {\bibinfo {author} {\bibfnamefont {D.}~\bibnamefont
  {Luo}}, \bibinfo {author} {\bibfnamefont {D.~D.}\ \bibnamefont {Dai}},\ and\
  \bibinfo {author} {\bibfnamefont {L.}~\bibnamefont {Fu}},\ }\href
  {https://doi.org/10.48550/arXiv.2406.17645} {\bibinfo {title} {Simulating
  moiré quantum matter with neural network}} (\bibinfo {year} {2024}),\
  \bibinfo {note} {arXiv:2406.17645 [cond-mat]}\BibitemShut {NoStop}%
\bibitem [{\citenamefont {Luo}\ \emph {et~al.}(2025)\citenamefont {Luo},
  \citenamefont {Zaklama},\ and\ \citenamefont {Fu}}]{luo_solving_2025}%
  \BibitemOpen
  \bibfield  {author} {\bibinfo {author} {\bibfnamefont {D.}~\bibnamefont
  {Luo}}, \bibinfo {author} {\bibfnamefont {T.}~\bibnamefont {Zaklama}},\ and\
  \bibinfo {author} {\bibfnamefont {L.}~\bibnamefont {Fu}},\ }\href
  {https://doi.org/10.48550/arXiv.2503.13585} {\bibinfo {title} {Solving
  fractional electron states in twisted {MoTe}$_2$ with deep neural networks}}
  (\bibinfo {year} {2025}),\ \bibinfo {note} {arXiv:2503.13585
  [cond-mat]}\BibitemShut {NoStop}%
\bibitem [{\citenamefont {Ruggeri}\ \emph {et~al.}(2018)\citenamefont
  {Ruggeri}, \citenamefont {Moroni},\ and\ \citenamefont
  {Holzmann}}]{2018_ruggeri_backflow}%
  \BibitemOpen
  \bibfield  {author} {\bibinfo {author} {\bibfnamefont {M.}~\bibnamefont
  {Ruggeri}}, \bibinfo {author} {\bibfnamefont {S.}~\bibnamefont {Moroni}},\
  and\ \bibinfo {author} {\bibfnamefont {M.}~\bibnamefont {Holzmann}},\
  }\bibfield  {title} {\bibinfo {title} {Nonlinear network description for
  many-body quantum systems in continuous space},\ }\href
  {https://doi.org/10.1103/PhysRevLett.120.205302} {\bibfield  {journal}
  {\bibinfo  {journal} {Phys. Rev. Lett.}\ }\textbf {\bibinfo {volume} {120}},\
  \bibinfo {pages} {205302} (\bibinfo {year} {2018})}\BibitemShut {NoStop}%
\bibitem [{\citenamefont {Linteau}\ \emph {et~al.}(2025)\citenamefont
  {Linteau}, \citenamefont {Pescia}, \citenamefont {Nys}, \citenamefont
  {Carleo},\ and\ \citenamefont {Holzmann}}]{2024_linteau_helium_4_2d}%
  \BibitemOpen
  \bibfield  {author} {\bibinfo {author} {\bibfnamefont {D.}~\bibnamefont
  {Linteau}}, \bibinfo {author} {\bibfnamefont {G.}~\bibnamefont {Pescia}},
  \bibinfo {author} {\bibfnamefont {J.}~\bibnamefont {Nys}}, \bibinfo {author}
  {\bibfnamefont {G.}~\bibnamefont {Carleo}},\ and\ \bibinfo {author}
  {\bibfnamefont {M.}~\bibnamefont {Holzmann}},\ }\bibfield  {title} {\bibinfo
  {title} {Phase diagram and crystal melting of helium-4 in two dimensions},\
  }\href {https://doi.org/10.1103/v1g7-m9k4} {\bibfield  {journal} {\bibinfo
  {journal} {Phys. Rev. Lett.}\ }\textbf {\bibinfo {volume} {134}},\ \bibinfo
  {pages} {246001} (\bibinfo {year} {2025})}\BibitemShut {NoStop}%
\bibitem [{\citenamefont {Ewald}(1921)}]{1921_ewald}%
  \BibitemOpen
  \bibfield  {author} {\bibinfo {author} {\bibfnamefont {P.~P.}\ \bibnamefont
  {Ewald}},\ }\bibfield  {title} {\bibinfo {title} {Die berechnung optischer
  und elektrostatischer gitterpotentiale},\ }\href
  {https://doi.org/https://doi.org/10.1002/andp.19213690304} {\bibfield
  {journal} {\bibinfo  {journal} {Annalen der Physik}\ }\textbf {\bibinfo
  {volume} {369}},\ \bibinfo {pages} {253} (\bibinfo {year}
  {1921})}\BibitemShut {NoStop}%
\bibitem [{\citenamefont {Taddei}\ \emph {et~al.}(2015)\citenamefont {Taddei},
  \citenamefont {Ruggeri}, \citenamefont {Moroni},\ and\ \citenamefont
  {Holzmann}}]{2015_taddei_iterative_backflow_fermi_liquids}%
  \BibitemOpen
  \bibfield  {author} {\bibinfo {author} {\bibfnamefont {M.}~\bibnamefont
  {Taddei}}, \bibinfo {author} {\bibfnamefont {M.}~\bibnamefont {Ruggeri}},
  \bibinfo {author} {\bibfnamefont {S.}~\bibnamefont {Moroni}},\ and\ \bibinfo
  {author} {\bibfnamefont {M.}~\bibnamefont {Holzmann}},\ }\bibfield  {title}
  {\bibinfo {title} {Iterative backflow renormalization procedure for many-body
  ground-state wave functions of strongly interacting normal fermi liquids},\
  }\href {https://doi.org/10.1103/PhysRevB.91.115106} {\bibfield  {journal}
  {\bibinfo  {journal} {Phys. Rev. B}\ }\textbf {\bibinfo {volume} {91}},\
  \bibinfo {pages} {115106} (\bibinfo {year} {2015})}\BibitemShut {NoStop}%
\bibitem [{\citenamefont {Ceperley}(1991)}]{Ceperley91}%
  \BibitemOpen
  \bibfield  {author} {\bibinfo {author} {\bibfnamefont {D.~M.}\ \bibnamefont
  {Ceperley}},\ }\bibfield  {title} {\bibinfo {title} {Fermion nodes},\ }\href
  {https://doi.org/10.1007/BF01030009} {\bibfield  {journal} {\bibinfo
  {journal} {J. Stat. Phys.}\ }\textbf {\bibinfo {volume} {63}},\ \bibinfo
  {pages} {1237} (\bibinfo {year} {1991})}\BibitemShut {NoStop}%
\bibitem [{\citenamefont {Luo}\ and\ \citenamefont {Clark}(2019)}]{Clark19}%
  \BibitemOpen
  \bibfield  {author} {\bibinfo {author} {\bibfnamefont {D.}~\bibnamefont
  {Luo}}\ and\ \bibinfo {author} {\bibfnamefont {B.~K.}\ \bibnamefont
  {Clark}},\ }\bibfield  {title} {\bibinfo {title} {Backflow transformations
  via neural networks for quantum many-body wave functions},\ }\href
  {https://doi.org/10.1103/PhysRevLett.122.226401} {\bibfield  {journal}
  {\bibinfo  {journal} {Phys. Rev. Lett.}\ }\textbf {\bibinfo {volume} {122}},\
  \bibinfo {pages} {226401} (\bibinfo {year} {2019})}\BibitemShut {NoStop}%
\bibitem [{\citenamefont {Nosanow}(1964)}]{1964_nosanow_crystalline_he3}%
  \BibitemOpen
  \bibfield  {author} {\bibinfo {author} {\bibfnamefont {L.~H.}\ \bibnamefont
  {Nosanow}},\ }\bibfield  {title} {\bibinfo {title} {Theory of crystalline
  {He}$^3$ at {0}\,$^\circ$k},\ }\href
  {https://doi.org/10.1103/PhysRevLett.13.270} {\bibfield  {journal} {\bibinfo
  {journal} {Phys. Rev. Lett.}\ }\textbf {\bibinfo {volume} {13}},\ \bibinfo
  {pages} {270} (\bibinfo {year} {1964})}\BibitemShut {NoStop}%
\bibitem [{\citenamefont {Hansen}\ and\ \citenamefont
  {Levesque}(1968)}]{1968_hansen_gs_he4_he3}%
  \BibitemOpen
  \bibfield  {author} {\bibinfo {author} {\bibfnamefont {J.-P.}\ \bibnamefont
  {Hansen}}\ and\ \bibinfo {author} {\bibfnamefont {D.}~\bibnamefont
  {Levesque}},\ }\bibfield  {title} {\bibinfo {title} {Ground state of solid
  helium-4 and -3},\ }\href {https://doi.org/10.1103/PhysRev.165.293}
  {\bibfield  {journal} {\bibinfo  {journal} {Phys. Rev.}\ }\textbf {\bibinfo
  {volume} {165}},\ \bibinfo {pages} {293} (\bibinfo {year}
  {1968})}\BibitemShut {NoStop}%
\bibitem [{\citenamefont {Pescia}\ \emph {et~al.}(2024)\citenamefont {Pescia},
  \citenamefont {Nys}, \citenamefont {Kim}, \citenamefont {Lovato},\ and\
  \citenamefont {Carleo}}]{2024_pescia_mpnn_electrongas}%
  \BibitemOpen
  \bibfield  {author} {\bibinfo {author} {\bibfnamefont {G.}~\bibnamefont
  {Pescia}}, \bibinfo {author} {\bibfnamefont {J.}~\bibnamefont {Nys}},
  \bibinfo {author} {\bibfnamefont {J.}~\bibnamefont {Kim}}, \bibinfo {author}
  {\bibfnamefont {A.}~\bibnamefont {Lovato}},\ and\ \bibinfo {author}
  {\bibfnamefont {G.}~\bibnamefont {Carleo}},\ }\bibfield  {title} {\bibinfo
  {title} {Message-passing neural quantum states for the homogeneous electron
  gas},\ }\href {https://doi.org/10.1103/PhysRevB.110.035108} {\bibfield
  {journal} {\bibinfo  {journal} {Phys. Rev. B}\ }\textbf {\bibinfo {volume}
  {110}},\ \bibinfo {pages} {035108} (\bibinfo {year} {2024})}\BibitemShut
  {NoStop}%
\bibitem [{\citenamefont {Holzmann}\ \emph {et~al.}(2016)\citenamefont
  {Holzmann}, \citenamefont {Clay}, \citenamefont {Morales}, \citenamefont
  {Tubman}, \citenamefont {Ceperley},\ and\ \citenamefont
  {Pierleoni}}]{2016_holzmann_finite_size_effects}%
  \BibitemOpen
  \bibfield  {author} {\bibinfo {author} {\bibfnamefont {M.}~\bibnamefont
  {Holzmann}}, \bibinfo {author} {\bibfnamefont {R.~C.}\ \bibnamefont {Clay}},
  \bibinfo {author} {\bibfnamefont {M.~A.}\ \bibnamefont {Morales}}, \bibinfo
  {author} {\bibfnamefont {N.~M.}\ \bibnamefont {Tubman}}, \bibinfo {author}
  {\bibfnamefont {D.~M.}\ \bibnamefont {Ceperley}},\ and\ \bibinfo {author}
  {\bibfnamefont {C.}~\bibnamefont {Pierleoni}},\ }\bibfield  {title} {\bibinfo
  {title} {Theory of finite size effects for electronic quantum monte carlo
  calculations of liquids and solids},\ }\href
  {https://doi.org/10.1103/PhysRevB.94.035126} {\bibfield  {journal} {\bibinfo
  {journal} {Phys. Rev. B}\ }\textbf {\bibinfo {volume} {94}},\ \bibinfo
  {pages} {035126} (\bibinfo {year} {2016})}\BibitemShut {NoStop}%
\bibitem [{\citenamefont {Lin}\ \emph {et~al.}(2001)\citenamefont {Lin},
  \citenamefont {Zong},\ and\ \citenamefont
  {Ceperley}}]{2001_lin_ceperley_tabc}%
  \BibitemOpen
  \bibfield  {author} {\bibinfo {author} {\bibfnamefont {C.}~\bibnamefont
  {Lin}}, \bibinfo {author} {\bibfnamefont {F.~H.}\ \bibnamefont {Zong}},\ and\
  \bibinfo {author} {\bibfnamefont {D.~M.}\ \bibnamefont {Ceperley}},\
  }\bibfield  {title} {\bibinfo {title} {Twist-averaged boundary conditions in
  continuum quantum monte carlo algorithms},\ }\href
  {https://doi.org/10.1103/PhysRevE.64.016702} {\bibfield  {journal} {\bibinfo
  {journal} {Phys. Rev. E}\ }\textbf {\bibinfo {volume} {64}},\ \bibinfo
  {pages} {016702} (\bibinfo {year} {2001})}\BibitemShut {NoStop}%
\bibitem [{\citenamefont {Morales}\ \emph {et~al.}(2014)\citenamefont
  {Morales}, \citenamefont {Clay}, \citenamefont {Pierleoni},\ and\
  \citenamefont {Ceperley}}]{Morales14}%
  \BibitemOpen
  \bibfield  {author} {\bibinfo {author} {\bibfnamefont {M.~A.}\ \bibnamefont
  {Morales}}, \bibinfo {author} {\bibfnamefont {R.}~\bibnamefont {Clay}},
  \bibinfo {author} {\bibfnamefont {C.}~\bibnamefont {Pierleoni}},\ and\
  \bibinfo {author} {\bibfnamefont {D.~M.}\ \bibnamefont {Ceperley}},\
  }\bibfield  {title} {\bibinfo {title} {First principles methods: A
  perspective from quantum monte carlo},\ }\href
  {https://doi.org/10.3390/e16010287} {\bibfield  {journal} {\bibinfo
  {journal} {Entropy}\ }\textbf {\bibinfo {volume} {16}},\ \bibinfo {pages}
  {287} (\bibinfo {year} {2014})}\BibitemShut {NoStop}%
\bibitem [{\citenamefont {Zong}\ \emph {et~al.}(2020)\citenamefont {Zong},
  \citenamefont {Wiebe},\ and\ \citenamefont
  {Ackland}}]{zong2020understanding}%
  \BibitemOpen
  \bibfield  {author} {\bibinfo {author} {\bibfnamefont {H.}~\bibnamefont
  {Zong}}, \bibinfo {author} {\bibfnamefont {H.}~\bibnamefont {Wiebe}},\ and\
  \bibinfo {author} {\bibfnamefont {G.~J.}\ \bibnamefont {Ackland}},\
  }\bibfield  {title} {\bibinfo {title} {Understanding high pressure molecular
  hydrogen with a hierarchical machine-learned potential},\ }\href
  {https://doi.org/https://doi.org/10.1038/s41467-020-18788-9} {\bibfield
  {journal} {\bibinfo  {journal} {Nature communications}\ }\textbf {\bibinfo
  {volume} {11}},\ \bibinfo {pages} {5014} (\bibinfo {year}
  {2020})}\BibitemShut {NoStop}%
\bibitem [{\citenamefont {Cheng}\ \emph {et~al.}(2020)\citenamefont {Cheng},
  \citenamefont {Mazzola}, \citenamefont {Pickard},\ and\ \citenamefont
  {Ceriotti}}]{Cheng2020}%
  \BibitemOpen
  \bibfield  {author} {\bibinfo {author} {\bibfnamefont {B.}~\bibnamefont
  {Cheng}}, \bibinfo {author} {\bibfnamefont {G.}~\bibnamefont {Mazzola}},
  \bibinfo {author} {\bibfnamefont {C.~J.}\ \bibnamefont {Pickard}},\ and\
  \bibinfo {author} {\bibfnamefont {M.}~\bibnamefont {Ceriotti}},\ }\bibfield
  {title} {\bibinfo {title} {{Evidence for supercritical behaviour of
  high-pressure liquid hydrogen}},\ }\href
  {https://doi.org/10.1038/s41586-020-2677-y} {\bibfield  {journal} {\bibinfo
  {journal} {Nature}\ }\textbf {\bibinfo {volume} {585}},\ \bibinfo {pages}
  {217} (\bibinfo {year} {2020})}\BibitemShut {NoStop}%
\bibitem [{\citenamefont {Karasiev}\ \emph {et~al.}(2021)\citenamefont
  {Karasiev}, \citenamefont {Hinz}, \citenamefont {Hu},\ and\ \citenamefont
  {Trickey}}]{karasiev2021liquid}%
  \BibitemOpen
  \bibfield  {author} {\bibinfo {author} {\bibfnamefont {V.~V.}\ \bibnamefont
  {Karasiev}}, \bibinfo {author} {\bibfnamefont {J.}~\bibnamefont {Hinz}},
  \bibinfo {author} {\bibfnamefont {S.}~\bibnamefont {Hu}},\ and\ \bibinfo
  {author} {\bibfnamefont {S.}~\bibnamefont {Trickey}},\ }\bibfield  {title}
  {\bibinfo {title} {On the liquid--liquid phase transition of dense
  hydrogen},\ }\href
  {https://doi.org/https://doi.org/10.1038/s41586-021-04078-x} {\bibfield
  {journal} {\bibinfo  {journal} {Nature}\ }\textbf {\bibinfo {volume} {600}},\
  \bibinfo {pages} {E12} (\bibinfo {year} {2021})}\BibitemShut {NoStop}%
\bibitem [{\citenamefont {Istas}\ \emph {et~al.}(2025)\citenamefont {Istas},
  \citenamefont {Jensen}, \citenamefont {Yang}, \citenamefont {Holzmann},
  \citenamefont {Pierleoni},\ and\ \citenamefont
  {Ceperley}}]{istas2024liquidliquidphasetransitionhydrogen}%
  \BibitemOpen
  \bibfield  {author} {\bibinfo {author} {\bibfnamefont {M.}~\bibnamefont
  {Istas}}, \bibinfo {author} {\bibfnamefont {S.}~\bibnamefont {Jensen}},
  \bibinfo {author} {\bibfnamefont {Y.}~\bibnamefont {Yang}}, \bibinfo {author}
  {\bibfnamefont {M.}~\bibnamefont {Holzmann}}, \bibinfo {author}
  {\bibfnamefont {C.}~\bibnamefont {Pierleoni}},\ and\ \bibinfo {author}
  {\bibfnamefont {D.~M.}\ \bibnamefont {Ceperley}},\ }\bibfield  {title}
  {\bibinfo {title} {Liquid-liquid phase transition of hydrogen and its
  critical point: Analysis from ab initio simulation and a machine-learned
  potential},\ }\href {https://doi.org/10.1103/PhysRevE.111.045307} {\bibfield
  {journal} {\bibinfo  {journal} {Phys. Rev. E}\ }\textbf {\bibinfo {volume}
  {111}},\ \bibinfo {pages} {045307} (\bibinfo {year} {2025})}\BibitemShut
  {NoStop}%
\bibitem [{\citenamefont {Tenti}\ \emph {et~al.}(2025)\citenamefont {Tenti},
  \citenamefont {J\"ackl}, \citenamefont {Nakano}, \citenamefont {Rupp},\ and\
  \citenamefont {Casula}}]{tenti2025hydrogen}%
  \BibitemOpen
  \bibfield  {author} {\bibinfo {author} {\bibfnamefont {G.}~\bibnamefont
  {Tenti}}, \bibinfo {author} {\bibfnamefont {B.}~\bibnamefont {J\"ackl}},
  \bibinfo {author} {\bibfnamefont {K.}~\bibnamefont {Nakano}}, \bibinfo
  {author} {\bibfnamefont {M.}~\bibnamefont {Rupp}},\ and\ \bibinfo {author}
  {\bibfnamefont {M.}~\bibnamefont {Casula}},\ }\bibfield  {title} {\bibinfo
  {title} {Hydrogen liquid-liquid transition from first principles and machine
  learning},\ }\href {https://doi.org/10.1103/pbrk-3zgd} {\bibfield  {journal}
  {\bibinfo  {journal} {Phys. Rev. B}\ }\textbf {\bibinfo {volume} {112}},\
  \bibinfo {pages} {104208} (\bibinfo {year} {2025})}\BibitemShut {NoStop}%
\bibitem [{\citenamefont {Clay}\ \emph {et~al.}(2016)\citenamefont {Clay},
  \citenamefont {Holzmann}, \citenamefont {Ceperley},\ and\ \citenamefont
  {Morales}}]{Clay2016}%
  \BibitemOpen
  \bibfield  {author} {\bibinfo {author} {\bibfnamefont {R.~C.}\ \bibnamefont
  {Clay}}, \bibinfo {author} {\bibfnamefont {M.}~\bibnamefont {Holzmann}},
  \bibinfo {author} {\bibfnamefont {D.~M.}\ \bibnamefont {Ceperley}},\ and\
  \bibinfo {author} {\bibfnamefont {M.~A.}\ \bibnamefont {Morales}},\
  }\bibfield  {title} {\bibinfo {title} {{Benchmarking density functionals for
  hydrogen-helium mixtures with quantum Monte Carlo: Energetics, pressures, and
  forces}},\ }\href {https://doi.org/10.1103/PhysRevB.93.035121} {\bibfield
  {journal} {\bibinfo  {journal} {Phys. Rev. B}\ }\textbf {\bibinfo {volume}
  {93}},\ \bibinfo {pages} {035121} (\bibinfo {year} {2016})}\BibitemShut
  {NoStop}%
\bibitem [{\citenamefont {Cozza}\ \emph {et~al.}(2026)\citenamefont {Cozza},
  \citenamefont {Nakano}, \citenamefont {Howard}, \citenamefont {Xie},
  \citenamefont {Helled},\ and\ \citenamefont {Mazzola}}]{cozza2025}%
  \BibitemOpen
  \bibfield  {author} {\bibinfo {author} {\bibfnamefont {C.}~\bibnamefont
  {Cozza}}, \bibinfo {author} {\bibfnamefont {K.}~\bibnamefont {Nakano}},
  \bibinfo {author} {\bibfnamefont {S.}~\bibnamefont {Howard}}, \bibinfo
  {author} {\bibfnamefont {H.}~\bibnamefont {Xie}}, \bibinfo {author}
  {\bibfnamefont {R.}~\bibnamefont {Helled}},\ and\ \bibinfo {author}
  {\bibfnamefont {G.}~\bibnamefont {Mazzola}},\ }\bibfield  {title} {\bibinfo
  {title} {Denser hydrogen inferred from first-principles simulations
  challenges {J}upiter's interior models},\ }\href
  {https://doi.org/10.1103/yrk6-ryps} {\bibfield  {journal} {\bibinfo
  {journal} {Phys. Rev. Res.}\ }\textbf {\bibinfo {volume} {8}},\ \bibinfo
  {pages} {013089} (\bibinfo {year} {2026})}\BibitemShut {NoStop}%
\bibitem [{\citenamefont {Yang}\ \emph {et~al.}(2022)\citenamefont {Yang},
  \citenamefont {Jensen}, \citenamefont {Ceperley}, \citenamefont {Kowalik},\
  and\ \citenamefont {Turk}}]{QMC-hamm}%
  \BibitemOpen
  \bibfield  {author} {\bibinfo {author} {\bibfnamefont {Y.}~\bibnamefont
  {Yang}}, \bibinfo {author} {\bibfnamefont {S.}~\bibnamefont {Jensen}},
  \bibinfo {author} {\bibfnamefont {D.}~\bibnamefont {Ceperley}}, \bibinfo
  {author} {\bibfnamefont {K.}~\bibnamefont {Kowalik}},\ and\ \bibinfo {author}
  {\bibfnamefont {M.}~\bibnamefont {Turk}},\ }\href
  {https://qmc-hamm.hub.yt/data.html} {\bibinfo {title} {Dense hydrogen
  diffusion monte carlo database}} (\bibinfo {year} {2022})\BibitemShut
  {NoStop}%
\bibitem [{\citenamefont {Rende}\ \emph {et~al.}(2025)\citenamefont {Rende},
  \citenamefont {Viteritti}, \citenamefont {Becca}, \citenamefont
  {Scardicchio}, \citenamefont {Laio},\ and\ \citenamefont
  {Carleo}}]{2025_rende_foundation_neural_network_quantum_states}%
  \BibitemOpen
  \bibfield  {author} {\bibinfo {author} {\bibfnamefont {R.}~\bibnamefont
  {Rende}}, \bibinfo {author} {\bibfnamefont {L.~L.}\ \bibnamefont
  {Viteritti}}, \bibinfo {author} {\bibfnamefont {F.}~\bibnamefont {Becca}},
  \bibinfo {author} {\bibfnamefont {A.}~\bibnamefont {Scardicchio}}, \bibinfo
  {author} {\bibfnamefont {A.}~\bibnamefont {Laio}},\ and\ \bibinfo {author}
  {\bibfnamefont {G.}~\bibnamefont {Carleo}},\ }\bibfield  {title} {\bibinfo
  {title} {Foundation neural-networks quantum states as a unified ansatz for
  multiple hamiltonians},\ }\href {https://doi.org/10.1038/s41467-025-62098-x}
  {\bibfield  {journal} {\bibinfo  {journal} {Nature Communications}\ }\textbf
  {\bibinfo {volume} {16}},\ \bibinfo {pages} {7213} (\bibinfo {year}
  {2025})}\BibitemShut {NoStop}%
\bibitem [{\citenamefont {Jones}\ and\ \citenamefont
  {Ceperley}(1996)}]{Jones96}%
  \BibitemOpen
  \bibfield  {author} {\bibinfo {author} {\bibfnamefont {M.~D.}\ \bibnamefont
  {Jones}}\ and\ \bibinfo {author} {\bibfnamefont {D.~M.}\ \bibnamefont
  {Ceperley}},\ }\bibfield  {title} {\bibinfo {title} {Crystallization of the
  one-component plasma at finite temperature},\ }\href
  {https://doi.org/10.1103/PhysRevLett.76.4572} {\bibfield  {journal} {\bibinfo
   {journal} {Phys. Rev. Lett.}\ }\textbf {\bibinfo {volume} {76}},\ \bibinfo
  {pages} {4572} (\bibinfo {year} {1996})}\BibitemShut {NoStop}%
\bibitem [{\citenamefont {Mon}\ \emph {et~al.}(1980)\citenamefont {Mon},
  \citenamefont {Chester},\ and\ \citenamefont
  {Ashcroft}}]{1980_mon_high_pressure_hydrogen_melting}%
  \BibitemOpen
  \bibfield  {author} {\bibinfo {author} {\bibfnamefont {K.~K.}\ \bibnamefont
  {Mon}}, \bibinfo {author} {\bibfnamefont {G.~V.}\ \bibnamefont {Chester}},\
  and\ \bibinfo {author} {\bibfnamefont {N.~W.}\ \bibnamefont {Ashcroft}},\
  }\bibfield  {title} {\bibinfo {title} {Simulation studies of a model of
  high-density metallic hydrogen},\ }\href
  {https://doi.org/10.1103/PhysRevB.21.2641} {\bibfield  {journal} {\bibinfo
  {journal} {Phys. Rev. B}\ }\textbf {\bibinfo {volume} {21}},\ \bibinfo
  {pages} {2641} (\bibinfo {year} {1980})}\BibitemShut {NoStop}%
\bibitem [{\citenamefont
  {Ceperley}(1989)}]{1989_ceperley_qmc_high_pressure_melting}%
  \BibitemOpen
  \bibfield  {author} {\bibinfo {author} {\bibfnamefont {D.~M.}\ \bibnamefont
  {Ceperley}},\ }\bibinfo {title} {Quantum monte carlo simulation of systems at
  high pressure},\ in\ \href@noop {} {\emph {\bibinfo {booktitle} {Simple
  Molecular Systems at Very High Density}}},\ \bibinfo {editor} {edited by\
  \bibinfo {editor} {\bibfnamefont {P.~L.}\ \bibnamefont {A.~Polian}}\ and\
  \bibinfo {editor} {\bibfnamefont {N.}~\bibnamefont {Boccara}}}\ (\bibinfo
  {publisher} {Plenum Press},\ \bibinfo {address} {New York},\ \bibinfo {year}
  {1989})\BibitemShut {NoStop}%
\bibitem [{\citenamefont {Militzer}\ and\ \citenamefont
  {Graham}(2006)}]{2006_militzer_melting_transition_hydrogen}%
  \BibitemOpen
  \bibfield  {author} {\bibinfo {author} {\bibfnamefont {B.}~\bibnamefont
  {Militzer}}\ and\ \bibinfo {author} {\bibfnamefont {R.~L.}\ \bibnamefont
  {Graham}},\ }\bibfield  {title} {\bibinfo {title} {Simulations of dense
  atomic hydrogen in the wigner crystal phase},\ }\href
  {https://doi.org/https://doi.org/10.1016/j.jpcs.2006.05.015} {\bibfield
  {journal} {\bibinfo  {journal} {Journal of Physics and Chemistry of Solids}\
  }\textbf {\bibinfo {volume} {67}},\ \bibinfo {pages} {2136} (\bibinfo {year}
  {2006})}\BibitemShut {NoStop}%
\bibitem [{\citenamefont {Carleo}\ \emph {et~al.}(2019)\citenamefont {Carleo},
  \citenamefont {Choo}, \citenamefont {Hofmann}, \citenamefont {Smith},
  \citenamefont {Westerhout}, \citenamefont {Alet}, \citenamefont {Davis},
  \citenamefont {Efthymiou}, \citenamefont {Glasser}, \citenamefont {Lin},
  \citenamefont {Mauri}, \citenamefont {Mazzola}, \citenamefont {Mendl},
  \citenamefont {{van Nieuwenburg}}, \citenamefont {O’Reilly}, \citenamefont
  {Théveniaut}, \citenamefont {Torlai}, \citenamefont {Vicentini},\ and\
  \citenamefont {Wietek}}]{2019_netket}%
  \BibitemOpen
  \bibfield  {author} {\bibinfo {author} {\bibfnamefont {G.}~\bibnamefont
  {Carleo}}, \bibinfo {author} {\bibfnamefont {K.}~\bibnamefont {Choo}},
  \bibinfo {author} {\bibfnamefont {D.}~\bibnamefont {Hofmann}}, \bibinfo
  {author} {\bibfnamefont {J.~E.}\ \bibnamefont {Smith}}, \bibinfo {author}
  {\bibfnamefont {T.}~\bibnamefont {Westerhout}}, \bibinfo {author}
  {\bibfnamefont {F.}~\bibnamefont {Alet}}, \bibinfo {author} {\bibfnamefont
  {E.~J.}\ \bibnamefont {Davis}}, \bibinfo {author} {\bibfnamefont
  {S.}~\bibnamefont {Efthymiou}}, \bibinfo {author} {\bibfnamefont
  {I.}~\bibnamefont {Glasser}}, \bibinfo {author} {\bibfnamefont {S.-H.}\
  \bibnamefont {Lin}}, \bibinfo {author} {\bibfnamefont {M.}~\bibnamefont
  {Mauri}}, \bibinfo {author} {\bibfnamefont {G.}~\bibnamefont {Mazzola}},
  \bibinfo {author} {\bibfnamefont {C.~B.}\ \bibnamefont {Mendl}}, \bibinfo
  {author} {\bibfnamefont {E.}~\bibnamefont {{van Nieuwenburg}}}, \bibinfo
  {author} {\bibfnamefont {O.}~\bibnamefont {O’Reilly}}, \bibinfo {author}
  {\bibfnamefont {H.}~\bibnamefont {Théveniaut}}, \bibinfo {author}
  {\bibfnamefont {G.}~\bibnamefont {Torlai}}, \bibinfo {author} {\bibfnamefont
  {F.}~\bibnamefont {Vicentini}},\ and\ \bibinfo {author} {\bibfnamefont
  {A.}~\bibnamefont {Wietek}},\ }\bibfield  {title} {\bibinfo {title} {Netket:
  A machine learning toolkit for many-body quantum systems},\ }\href
  {https://doi.org/https://doi.org/10.1016/j.softx.2019.100311} {\bibfield
  {journal} {\bibinfo  {journal} {SoftwareX}\ }\textbf {\bibinfo {volume}
  {10}},\ \bibinfo {pages} {100311} (\bibinfo {year} {2019})}\BibitemShut
  {NoStop}%
\bibitem [{\citenamefont {Vicentini}\ \emph {et~al.}(2022)\citenamefont
  {Vicentini}, \citenamefont {Hofmann}, \citenamefont {Szabó}, \citenamefont
  {Wu}, \citenamefont {Roth}, \citenamefont {Giuliani}, \citenamefont {Pescia},
  \citenamefont {Nys}, \citenamefont {Vargas-Calderón}, \citenamefont
  {Astrakhantsev},\ and\ \citenamefont {Carleo}}]{2022_netket_3}%
  \BibitemOpen
  \bibfield  {author} {\bibinfo {author} {\bibfnamefont {F.}~\bibnamefont
  {Vicentini}}, \bibinfo {author} {\bibfnamefont {D.}~\bibnamefont {Hofmann}},
  \bibinfo {author} {\bibfnamefont {A.}~\bibnamefont {Szabó}}, \bibinfo
  {author} {\bibfnamefont {D.}~\bibnamefont {Wu}}, \bibinfo {author}
  {\bibfnamefont {C.}~\bibnamefont {Roth}}, \bibinfo {author} {\bibfnamefont
  {C.}~\bibnamefont {Giuliani}}, \bibinfo {author} {\bibfnamefont
  {G.}~\bibnamefont {Pescia}}, \bibinfo {author} {\bibfnamefont
  {J.}~\bibnamefont {Nys}}, \bibinfo {author} {\bibfnamefont {V.}~\bibnamefont
  {Vargas-Calderón}}, \bibinfo {author} {\bibfnamefont {N.}~\bibnamefont
  {Astrakhantsev}},\ and\ \bibinfo {author} {\bibfnamefont {G.}~\bibnamefont
  {Carleo}},\ }\bibfield  {title} {\bibinfo {title} {Netket 3: Machine learning
  toolbox for many-body quantum systems},\ }\href
  {https://doi.org/10.21468/SciPostPhysCodeb.7} {\bibfield  {journal} {\bibinfo
   {journal} {SciPost Physics Codebases}\ ,\ \bibinfo {eid} {7}} (\bibinfo
  {year} {2022})}\BibitemShut {NoStop}%
\bibitem [{\citenamefont {Bradbury}\ \emph {et~al.}(2025)\citenamefont
  {Bradbury}, \citenamefont {Frostig}, \citenamefont {Hawkins}, \citenamefont
  {Johnson}, \citenamefont {Leary}, \citenamefont {Maclaurin}, \citenamefont
  {Necula}, \citenamefont {Paszke}, \citenamefont {Vander{P}las}, \citenamefont
  {Wanderman-{M}ilne},\ and\ \citenamefont {Zhang}}]{2018_jax}%
  \BibitemOpen
  \bibfield  {author} {\bibinfo {author} {\bibfnamefont {J.}~\bibnamefont
  {Bradbury}}, \bibinfo {author} {\bibfnamefont {R.}~\bibnamefont {Frostig}},
  \bibinfo {author} {\bibfnamefont {P.}~\bibnamefont {Hawkins}}, \bibinfo
  {author} {\bibfnamefont {M.~J.}\ \bibnamefont {Johnson}}, \bibinfo {author}
  {\bibfnamefont {C.}~\bibnamefont {Leary}}, \bibinfo {author} {\bibfnamefont
  {D.}~\bibnamefont {Maclaurin}}, \bibinfo {author} {\bibfnamefont
  {G.}~\bibnamefont {Necula}}, \bibinfo {author} {\bibfnamefont
  {A.}~\bibnamefont {Paszke}}, \bibinfo {author} {\bibfnamefont
  {J.}~\bibnamefont {Vander{P}las}}, \bibinfo {author} {\bibfnamefont
  {S.}~\bibnamefont {Wanderman-{M}ilne}},\ and\ \bibinfo {author}
  {\bibfnamefont {Q.}~\bibnamefont {Zhang}},\ }\href@noop {} {\bibinfo {title}
  {{JAX}: composable transformations of {P}ython+{N}um{P}y programs}},\
  \bibinfo {howpublished} {\url{http://github.com/jax-ml/jax}} (\bibinfo {year}
  {2025})\BibitemShut {NoStop}%
\bibitem [{\citenamefont {Häfner}\ and\ \citenamefont
  {Vicentini}(2021)}]{2021_hafner_vicentini_mpi4jax}%
  \BibitemOpen
  \bibfield  {author} {\bibinfo {author} {\bibfnamefont {D.}~\bibnamefont
  {Häfner}}\ and\ \bibinfo {author} {\bibfnamefont {F.}~\bibnamefont
  {Vicentini}},\ }\bibfield  {title} {\bibinfo {title} {mpi4jax: Zero-copy
  {MPI} communication of {JAX} arrays},\ }\href
  {https://doi.org/10.21105/joss.03419} {\bibfield  {journal} {\bibinfo
  {journal} {Journal of Open Source Software}\ }\textbf {\bibinfo {volume}
  {6}},\ \bibinfo {pages} {3419} (\bibinfo {year} {2021})}\BibitemShut
  {NoStop}%
\bibitem [{\citenamefont {Linteau}(2026)}]{github_repo}%
  \BibitemOpen
  \bibfield  {author} {\bibinfo {author} {\bibfnamefont {D.}~\bibnamefont
  {Linteau}},\ }\href@noop {} {\bibinfo {title} {Metallic hydrogen}},\ \bibinfo
  {howpublished} {\url{https://github.com/cqsl/metallic-hydrogen}} (\bibinfo
  {year} {2026})\BibitemShut {NoStop}%
\bibitem [{\citenamefont {Li}\ \emph {et~al.}(2024)\citenamefont {Li},
  \citenamefont {Ye}, \citenamefont {Jiang}, \citenamefont {Wen}, \citenamefont
  {Wang}, \citenamefont {Li}, \citenamefont {Li}, \citenamefont {He},
  \citenamefont {Chen}, \citenamefont {Ren},\ and\ \citenamefont
  {Wang}}]{2024_li_forward_laplacian}%
  \BibitemOpen
  \bibfield  {author} {\bibinfo {author} {\bibfnamefont {R.}~\bibnamefont
  {Li}}, \bibinfo {author} {\bibfnamefont {H.}~\bibnamefont {Ye}}, \bibinfo
  {author} {\bibfnamefont {D.}~\bibnamefont {Jiang}}, \bibinfo {author}
  {\bibfnamefont {X.}~\bibnamefont {Wen}}, \bibinfo {author} {\bibfnamefont
  {C.}~\bibnamefont {Wang}}, \bibinfo {author} {\bibfnamefont {Z.}~\bibnamefont
  {Li}}, \bibinfo {author} {\bibfnamefont {X.}~\bibnamefont {Li}}, \bibinfo
  {author} {\bibfnamefont {D.}~\bibnamefont {He}}, \bibinfo {author}
  {\bibfnamefont {J.}~\bibnamefont {Chen}}, \bibinfo {author} {\bibfnamefont
  {W.}~\bibnamefont {Ren}},\ and\ \bibinfo {author} {\bibfnamefont
  {L.}~\bibnamefont {Wang}},\ }\bibfield  {title} {\bibinfo {title} {A
  computational framework for neural network-based variational monte carlo with
  forward laplacian},\ }\href {https://doi.org/10.1038/s42256-024-00794-x}
  {\bibfield  {journal} {\bibinfo  {journal} {Nature Machine Intelligence}\
  }\textbf {\bibinfo {volume} {6}},\ \bibinfo {pages} {209} (\bibinfo {year}
  {2024})}\BibitemShut {NoStop}%
\bibitem [{\citenamefont {Gao}\ \emph {et~al.}(2025)\citenamefont {Gao},
  \citenamefont {K{\"o}hler},\ and\ \citenamefont {Foster}}]{2023_gao_folx}%
  \BibitemOpen
  \bibfield  {author} {\bibinfo {author} {\bibfnamefont {N.}~\bibnamefont
  {Gao}}, \bibinfo {author} {\bibfnamefont {J.}~\bibnamefont {K{\"o}hler}},\
  and\ \bibinfo {author} {\bibfnamefont {A.}~\bibnamefont {Foster}},\
  }\href@noop {} {\bibinfo {title} {folx - forward laplacian for jax}},\
  \bibinfo {howpublished} {\url{http://github.com/microsoft/folx}} (\bibinfo
  {year} {2025})\BibitemShut {NoStop}%
\bibitem [{\citenamefont {Chen}(2024)}]{2024_chen_fwdlap}%
  \BibitemOpen
  \bibfield  {author} {\bibinfo {author} {\bibfnamefont {Y.}~\bibnamefont
  {Chen}},\ }\href@noop {} {\bibinfo {title} {Poor man's forward laplacian
  (using jax tracer!)}},\ \bibinfo {howpublished}
  {\url{https://github.com/y1xiaoc/fwdlap}} (\bibinfo {year}
  {2024})\BibitemShut {NoStop}%
\bibitem [{\citenamefont {Kim}\ \emph {et~al.}(2024)\citenamefont {Kim},
  \citenamefont {Pescia}, \citenamefont {Fore}, \citenamefont {Nys},
  \citenamefont {Carleo}, \citenamefont {Gandolfi}, \citenamefont
  {Hjorth-Jensen},\ and\ \citenamefont
  {Lovato}}]{2024_kim_ultra_cold_fermi_gases}%
  \BibitemOpen
  \bibfield  {author} {\bibinfo {author} {\bibfnamefont {J.}~\bibnamefont
  {Kim}}, \bibinfo {author} {\bibfnamefont {G.}~\bibnamefont {Pescia}},
  \bibinfo {author} {\bibfnamefont {B.}~\bibnamefont {Fore}}, \bibinfo {author}
  {\bibfnamefont {J.}~\bibnamefont {Nys}}, \bibinfo {author} {\bibfnamefont
  {G.}~\bibnamefont {Carleo}}, \bibinfo {author} {\bibfnamefont
  {S.}~\bibnamefont {Gandolfi}}, \bibinfo {author} {\bibfnamefont
  {M.}~\bibnamefont {Hjorth-Jensen}},\ and\ \bibinfo {author} {\bibfnamefont
  {A.}~\bibnamefont {Lovato}},\ }\bibfield  {title} {\bibinfo {title}
  {Neural-network quantum states for ultra-cold fermi gases},\ }\href
  {https://doi.org/10.1038/s42005-024-01613-w} {\bibfield  {journal} {\bibinfo
  {journal} {Communications Physics}\ }\textbf {\bibinfo {volume} {7}},\
  \bibinfo {pages} {148} (\bibinfo {year} {2024})}\BibitemShut {NoStop}%
\bibitem [{\citenamefont {Nys}\ \emph {et~al.}(2024)\citenamefont {Nys},
  \citenamefont {Pescia}, \citenamefont {Sinibaldi},\ and\ \citenamefont
  {Carleo}}]{2024_nys_tdvmc}%
  \BibitemOpen
  \bibfield  {author} {\bibinfo {author} {\bibfnamefont {J.}~\bibnamefont
  {Nys}}, \bibinfo {author} {\bibfnamefont {G.}~\bibnamefont {Pescia}},
  \bibinfo {author} {\bibfnamefont {A.}~\bibnamefont {Sinibaldi}},\ and\
  \bibinfo {author} {\bibfnamefont {G.}~\bibnamefont {Carleo}},\ }\bibfield
  {title} {\bibinfo {title} {Ab-initio variational wave functions for the
  time-dependent many-electron {S}chr{\"o}dinger equation},\ }\href@noop {}
  {\bibfield  {journal} {\bibinfo  {journal} {Nature Communications}\ }\textbf
  {\bibinfo {volume} {15}},\ \bibinfo {pages} {9404} (\bibinfo {year}
  {2024})}\BibitemShut {NoStop}%
\bibitem [{\citenamefont {Pierleoni}\ \emph
  {et~al.}(2008{\natexlab{b}})\citenamefont {Pierleoni}, \citenamefont
  {Delaney}, \citenamefont {Morales}, \citenamefont {Ceperley},\ and\
  \citenamefont {Holzmann}}]{2007_pierleoni_ceimc_hydrogen}%
  \BibitemOpen
  \bibfield  {author} {\bibinfo {author} {\bibfnamefont {C.}~\bibnamefont
  {Pierleoni}}, \bibinfo {author} {\bibfnamefont {K.~T.}\ \bibnamefont
  {Delaney}}, \bibinfo {author} {\bibfnamefont {M.~A.}\ \bibnamefont
  {Morales}}, \bibinfo {author} {\bibfnamefont {D.~M.}\ \bibnamefont
  {Ceperley}},\ and\ \bibinfo {author} {\bibfnamefont {M.}~\bibnamefont
  {Holzmann}},\ }\bibfield  {title} {\bibinfo {title} {Progress in coupled
  electron-ion monte carlo simulations of high-pressure hydrogen},\ }in\ \href
  {https://doi.org/10.1142/9789812779885_0029} {\emph {\bibinfo {booktitle}
  {Quantum Simulations of Complex Many-Body Systems: From Theory to
  Algorithms}}},\ \bibinfo {series} {Advances in Quantum Many-Body Theory},
  Vol.~\bibinfo {volume} {10},\ \bibinfo {editor} {edited by\ \bibinfo {editor}
  {\bibfnamefont {J.}~\bibnamefont {Grotendorst}}, \bibinfo {editor}
  {\bibfnamefont {D.}~\bibnamefont {Marx}},\ and\ \bibinfo {editor}
  {\bibfnamefont {A.}~\bibnamefont {Muramatsu}}}\ (\bibinfo  {publisher} {World
  Scientific Publishing},\ \bibinfo {address} {Singapore},\ \bibinfo {year}
  {2008})\ pp.\ \bibinfo {pages} {217--232}\BibitemShut {NoStop}%
\bibitem [{\citenamefont {Monkhorst}\ and\ \citenamefont
  {Pack}(1976)}]{1976_monkhorst_pack_grid}%
  \BibitemOpen
  \bibfield  {author} {\bibinfo {author} {\bibfnamefont {H.~J.}\ \bibnamefont
  {Monkhorst}}\ and\ \bibinfo {author} {\bibfnamefont {J.~D.}\ \bibnamefont
  {Pack}},\ }\bibfield  {title} {\bibinfo {title} {Special points for
  brillouin-zone integrations},\ }\href
  {https://doi.org/10.1103/PhysRevB.13.5188} {\bibfield  {journal} {\bibinfo
  {journal} {Phys. Rev. B}\ }\textbf {\bibinfo {volume} {13}},\ \bibinfo
  {pages} {5188} (\bibinfo {year} {1976})}\BibitemShut {NoStop}%
\bibitem [{\citenamefont {Sorella}(1998)}]{1998_sorella_sr}%
  \BibitemOpen
  \bibfield  {author} {\bibinfo {author} {\bibfnamefont {S.}~\bibnamefont
  {Sorella}},\ }\bibfield  {title} {\bibinfo {title} {Green function monte
  carlo with stochastic reconfiguration},\ }\href
  {https://doi.org/10.1103/PhysRevLett.80.4558} {\bibfield  {journal} {\bibinfo
   {journal} {Phys. Rev. Lett.}\ }\textbf {\bibinfo {volume} {80}},\ \bibinfo
  {pages} {4558} (\bibinfo {year} {1998})}\BibitemShut {NoStop}%
\bibitem [{\citenamefont {Sorella}(2005)}]{2005_sorella_sr}%
  \BibitemOpen
  \bibfield  {author} {\bibinfo {author} {\bibfnamefont {S.}~\bibnamefont
  {Sorella}},\ }\bibfield  {title} {\bibinfo {title} {Wave function
  optimization in the variational monte carlo method},\ }\href
  {https://doi.org/10.1103/PhysRevB.71.241103} {\bibfield  {journal} {\bibinfo
  {journal} {Phys. Rev. B}\ }\textbf {\bibinfo {volume} {71}},\ \bibinfo
  {pages} {241103} (\bibinfo {year} {2005})}\BibitemShut {NoStop}%
\bibitem [{\citenamefont {Becca}\ and\ \citenamefont
  {Sorella}(2017)}]{2017_becca_sorella_book}%
  \BibitemOpen
  \bibfield  {author} {\bibinfo {author} {\bibfnamefont {F.}~\bibnamefont
  {Becca}}\ and\ \bibinfo {author} {\bibfnamefont {S.}~\bibnamefont
  {Sorella}},\ }\href@noop {} {\emph {\bibinfo {title} {Quantum Monte Carlo
  Approaches for Correlated Systems}}}\ (\bibinfo  {publisher} {Cambridge
  University Press},\ \bibinfo {address} {Cambridge, UK},\ \bibinfo {year}
  {2017})\BibitemShut {NoStop}%
\end{thebibliography}%

\clearpage
\onecolumngrid

\appendix

\section{Message-passing neural network (MPNN) implementation} \label{appendix:mpnn_implementation}

The implementation used in this work is detailed in Algorithm 1. Other recent MPNN implementations for NQS were presented for instance in \cite{2024_pescia_mpnn_electrongas,2024_kim_ultra_cold_fermi_gases,2024_nys_tdvmc,2024_linteau_helium_4_2d}.

\vspace{0.5cm}
\hrule
\vspace{-0.2cm}
\begin{center}
    \textbf{Algorithm 1} MPNN
\end{center}
\vspace{-0.2cm}
\hrule
\vspace{0.2cm}
\textbf{Data--}
 Given a Monte-Carlo configuration $((\mathbf{r}_1,s_1), \hdots, (\mathbf{r}_N,s_N), \mathbf{R}_1, \hdots, \mathbf{R}_N)$, where $s_i$ is an electron spin label, the following electron-electron and electron-proton input tensors can be constructed:
\begin{align}
    \mathbf{I}_{ij} &= [\sin(2\pi \mathbf{r}_{ij}/L), \cos(2\pi \mathbf{r}_{ij}/L), |\sin(\pi \mathbf{r}_{ij}/L)|, s_i s_j] \in \mathbb{R}^{2d+2}, \\
    \mathbf{I}_{iI} &= [\sin(2\pi \mathbf{r}_{iI}/L), \cos(2\pi \mathbf{r}_{iI}/L), |\sin(\pi \mathbf{r}_{iI}/L)|] \in \mathbb{R}^{2d+1},
\end{align}
where $\mathbf{r}_{ij} = \mathbf{r}_i - \mathbf{r}_j$, $\mathbf{r}_{iI} = \mathbf{r}_i - \mathbf{R}_I$ and the square brackets correspond to the concatenation operation.
The following hyperparameters have to be fixed by the user: the hidden dimension $h$, the attention dimension $a$, the MLP output dimension $e$, and the number of message-passing iteration(s) $b$. \\

\noindent \textit{(The vector or matrix dimension is shown below as a comment on the right.)} \\
\noindent \textbf{Initialization--} instantiate ``hidden'' embedding parameters $ \mathbf{h}_i^{(0)}, \mathbf{H}_{ij}^{(0)}, \mathbf{H}_{iI}^{(0)}$ \hfill $\triangleright \ h$ \\
\noindent Define vertex states $\mathbf{y}_i^{(0)} = \mathbf{h}_i^{(0)}$ \hfill $\triangleright \ h$ \\
\noindent Define edge states $\mathbf{Y}_{ij}^{(0)} = [\mathbf{I}_{ij}, \mathbf{H}_{ij}^{(0)}]$, $\mathbf{Y}_{iI}^{(0)} = [\mathbf{I}_{iI}, \mathbf{H}_{iI}^{(0)}]$ \hfill $\triangleright \ 2d+2+h, \ 2d+1+h$ \\
\noindent \textbf{for} $0 \le \ell \le b$ \textbf{do} \\
\indent \textit{Electron-proton edge update}: \\
\indent Initialize query and key matrices $Q_{ep}^{(\ell)}, \ K_{ep}^{(\ell)}$ \hfill $\triangleright \ \mathrm{dim}(\mathbf{Y}_{iI}^{(\ell)}) \times a$ \\
\indent Compute queries and keys $\mathbf{Q}_{iI}^{(\ell)} = Q_{ep}^{(\ell)} \mathbf{Y}_{iI}^{(\ell)}, \ \ \mathbf{K}_{iI}^{(\ell)} = K_{ep}^{(\ell)} \mathbf{Y}_{iI}^{(\ell)}$ \hfill $\triangleright \ a$ \\
\indent Compute attention weights $\mathbf{W}_{iI}^{(\ell)} = \mathrm{MLP}_{ep,1}^{(\ell)} ( \mathbf{Q}^{(\ell)} \mathbf{K}^{(\ell)} / \sqrt{a} )_{iI}$ \hfill $\triangleright \ e$ \\
\indent Compute messages $\mathbf{M}_{iI}^{(\ell)} = \mathbf{W}_{iI}^{(\ell)} \odot \mathrm{MLP}_{ep,2}^{(\ell)} (\mathbf{Y}_{iI}^{(\ell)})$ \hfill $\triangleright \ e$ \\
\indent Update edge hidden states $\mathbf{H}_{iI}^{(\ell+1)} = \mathrm{MLP}_{ep,3}^{(\ell)} ([ \mathbf{Y}_{iI}^{(\ell)}, \mathbf{M}_{iI}^{(\ell)}])$ \hfill $\triangleright \ e$ \\
\vspace{0.5cm}
\indent Update edge states $\mathbf{Y}_{iI}^{(\ell+1)} = [ \mathbf{I}_{iI}, \mathbf{H}_{iI}^{(\ell+1)} ]$ \hfill $\triangleright \ 2d+1+e$ \\
\indent \textit{Electron-electron edge and vertex update}: \\
\indent Initialize query and key matrices $Q_{ee}^{(\ell)}, \ K_{ee}^{(\ell)}$ \hfill $\triangleright \ \mathrm{dim}(\mathbf{Y}_{ij}^{(\ell)}) \times a$ \\
\indent Compute queries and keys $\mathbf{Q}_{ij}^{(\ell)} = Q_{ee}^{(\ell)} \mathbf{Y}_{ij}^{(\ell)}, \ \ \mathbf{K}_{ij}^{(\ell)} = K_{ee}^{(\ell)} \mathbf{Y}_{ij}^{(\ell)}$ \hfill $\triangleright \ a$ \\
\indent Compute attention weights $\mathbf{W}_{ij}^{(\ell)} = \mathrm{MLP}_{ee,1}^{(\ell)} ( \mathbf{Q}^{(\ell)} \mathbf{K}^{(\ell)} / \sqrt{a} )_{ij}$ \hfill $\triangleright \ e$ \\
\indent Compute messages $\mathbf{M}_{ij}^{(\ell)} = \mathbf{W}_{ij}^{(\ell)} \odot \mathrm{MLP}_{ee,2}^{(\ell)} (\mathbf{Y}_{ij}^{(\ell)})$ \hfill $\triangleright \ e$ \\
\indent Compute vertex cross-species contributions: $\mathbf{C}_i^{(\ell)} = \sum_I \mathrm{MLP}_{ee,3}(\mathbf{Y}_{iI}^{(\ell+1)})$ \hfill $\triangleright \ e$ \\
\indent Compute edge cross-species contributions: $\mathbf{C}_{ij}^{(\ell)} = \sum_I \mathrm{MLP}_{ee,4}(\mathbf{Y}_{iI}^{(\ell+1)}) \odot \mathrm{MLP}_{ee,5}(\mathbf{Y}_{jI}^{(\ell+1)})$ \hfill $\triangleright \ e$ \\
\indent Update vertex hidden states $\mathbf{h}_i^{(\ell+1)} = \mathrm{MLP}_{ee,6}^{(\ell)} ([ \mathbf{h}_i^{(\ell)} , \sum_j \mathbf{M}_{ij}^{(\ell)}, \mathbf{C}_i^{(\ell)} ])$ \hfill $\triangleright \ e$ \\
\indent Update edge hidden states $\mathbf{H}_{ij}^{(\ell+1)} = \mathrm{MLP}_{ee,7}^{(\ell)} ([ \mathbf{Y}_{ij}^{(\ell)} , \mathbf{M}_{ij}^{(\ell)}, \mathbf{C}_{ij}^{(\ell)}])$ \hfill $\triangleright \ e$ \\
\indent Update vertex states $\mathbf{y}_i^{(\ell+1)} = \mathbf{h}_i^{(\ell+1)}$ \hfill $\triangleright \ e$ \\
\indent Update edge states $\mathbf{Y}_{ij}^{(\ell+1)} = [ \mathbf{I}_{ij}, \mathbf{H}_{ij}^{(\ell+1)} ]$ \hfill $\triangleright \ 2d+2+e$ \\
\noindent \textbf{end for} \\
\noindent \textbf{return} final vertex and edge states $\mathbf{y}_i^{(b)}, \mathbf{Y}_{ij}^{(b)}, \mathbf{Y}_{iI}^{(b)}$

\vspace{0.3cm}
\hrule

\clearpage
\section{Energy comparison for various crystal structures} \label{appendix:energy_comparison}

In \cref{table:static_hydrogen_rs=1_rs=1.4_pbc_and_tabc}, we compare twist-averaged (TABC) calculations for static protons with different crystal structures, as well as different $N$ and $r_s$, with the best reptation Monte Carlo (RMC) results from Ref.~\cite{2007_pierleoni_ceimc_hydrogen,Pierleoni08}, obtained either by using backflow augmented metallic (BF-PW), DFT (BF-LDA), or optimized independent particle potential (BF-IPP) orbitals. All TABC results are based on a $6 \times 6 \times 6$ twist grid.
The NQS results were obtained with a $6 \times 6 \times 6$ Monkhorst-Pack grid \cite{1976_monkhorst_pack_grid},  providing twist grid convergence within $\sim 0.5$ mHa.
We further provide PBC energies to facilitate the comparison for future work.
We find that the NQS generally performs as good or better than the RMC wave functions, except for the diamond lattice where NQS energies are slightly higher.

\begin{table}[h]
    \centering

    \begin{tabular}{|c|c|c|c|c|c|c|c|}
    \hline
    \multicolumn{2}{|c|}{System} & \multirow{2}{*}{\begin{tabular}{c} Boundary \\ condition \end{tabular}} & \multirow{2}{*}{Wave function} & \multicolumn{2}{c|}{$r_s=1$} & \multicolumn{2}{c|}{$r_s=1.4$} \\
    \cline{1-2}\cline{5-8}
    $N$ & Lattice &  &  & $E/N$ & $\sigma^2/N$ & $E/N$ & $\sigma^2/N$ \\
    \hline\hline
    \multirow[c]{4}{*}{54} & \multirow[c]{4}{*}{BCC} & PBC & \textbf{NQS (VMC)} & \textbf{-0.42232(2)} & \textbf{0.0101(1)} & \textbf{-0.55003(3)} & \textbf{0.00392(5)} \\
    \cline{3-8}
    & & \multirow[c]{3}{*}{TABC} & BF-PW (RMC) & -0.3721(1) & 0.0182(7) & - & - \\
    & & & BF-IPP (RMC) & - & - & -0.5228(1) & 0.01413(7) \\
    & & & \textbf{NQS (VMC)} & \textbf{-0.37519(6)} & \textbf{0.01298(3)} & \textbf{-0.52613(3)} & \textbf{0.00656(2)} \\
    \hline\hline
    \multirow[c]{4}{*}{32} & \multirow[c]{4}{*}{FCC} & PBC & \textbf{NQS (VMC)} & \textbf{-0.35927(1)} & \textbf{0.00983(3)} & \textbf{-0.51956(7)} & \textbf{0.00365(2)} \\
    \cline{3-8}
    & & \multirow[c]{3}{*}{TABC} & BF-PW (RMC)& -0.3792(1) & 0.01543(4) & - & - \\
    & & & BF-IPP (RMC) & - & - & -0.5280(1) & 0.01352(5) \\
    & & & \textbf{NQS (VMC)} & \textbf{-0.3820(1)} & \textbf{0.01276(5)} & \textbf{-0.53083(5)} & \textbf{0.00604(1)} \\
    \hline\hline
    \multirow[c]{4}{*}{64} & \multirow[c]{4}{*}{DIAM} & PBC & \textbf{NQS (VMC)} & \textbf{-0.35473(7)} & \textbf{0.0176(1)} & \textbf{-0.51887(7)} & \textbf{0.01027(4)} \\
    \cline{3-8}
    & & \multirow[c]{3}{*}{TABC} & \textbf{BF-LDA (RMC)} & \textbf{-0.3635(1)} & \textbf{0.0406(1)} & - & - \\
    & & & \textbf{BF-IPP (RMC)} & - & - & \textbf{-0.5346(1)} & \textbf{0.01740(7)} \\
    & & & NQS (VMC)& -0.35993(4) & 0.01225(5) & -0.52826(4) & 0.00684(1) \\
    \hline\hline
    \multirow[c]{3}{*}{8} & \multirow[c]{3}{*}{DIAM} & PBC & \textbf{NQS (VMC)} & \textbf{-0.10057(7)} & \textbf{0.00821(3)} & - & - \\
    \cline{3-8}
    & & \multirow[c]{2}{*}{TABC} & BF-PW (RMC) & -0.41368(6) & 0.01032(2) & - & - \\
    & & & \textbf{NQS (VMC)} & \textbf{-0.4223(3)} & \textbf{0.01026(4)} & - & - \\
    \hline\hline
    \multirow[c]{2}{*}{8} & \multirow[c]{2}{*}{SC} & PBC & \textbf{NQS (VMC)} & \textbf{-0.25179(3)} & \textbf{0.01051(8)} & - & - \\
    \cline{3-8}
    & & TABC & \textbf{NQS (VMC)} & \textbf{-0.4212(3)} & \textbf{0.01474(4)} & - & - \\
    \hline
    \end{tabular}
    \caption{
   Hydrogen with static protons for various crystal structures. The lattices considered include body-centered cubic (BCC), face-centered cubic (FCC), diamond (DIAM) and simple cubic (SC).
    The following acronyms are used: ``BF-PW'' corresponds to a metallic wave function using backflow plane wave orbitals, ``BF-LDA'' denotes
    LDA-DFT orbitals with backflow, ``BF-IPP'' stands for Independent Particle Potential (IPP) with backflow where the plane wave coefficients of the orbitals are obtained by solving an eigenvalue problem with an effective electron-ion potential.
    Those reference energies, taken from Ref.~\cite{2007_pierleoni_ceimc_hydrogen,Pierleoni08}, were obtained with RMC and a $6 \times 6 \times 6$ twist grid. Our VMC results of the
    NQS wave function are obtained using a $6 \times 6 \times 6$ Monkhorst-Pack \cite{1976_monkhorst_pack_grid} twist grid  when applying TABC, values at the $\Gamma$-Point (PBC) are provided as reference.
    Boldface entries indicate the lowest energy value within each set of compared results.
    }
    \label{table:static_hydrogen_rs=1_rs=1.4_pbc_and_tabc}
\end{table}

\section{Global optimization of a universal wave function} \label{appendix:global_optimization}

In developing a foundation model for electronic structure, the objective is to optimize a single set of variational parameters, $\boldsymbol{\theta}$, so that the electronic ground state wave function $\Psi_{\boldsymbol{\theta}}(\mathbf{R})$
accurately describes a set of electronic Hamiltonians $\{H_{e}(\mathbf{R})\}_\mathbf{R}$ (with $H_{e}(\mathbf{R})$ defined in \cref{eq:HBO}) associated with different static nuclear configurations $\{\mathbf{R}\}$, with $\mathbf{R} = \{ \mathbf{R}_1, \hdots, \mathbf{R}_N\}$. 
In the most general form, the associated global variational energy $E_\mathrm{global}(\boldsymbol{\theta})$ is given by
\begin{equation} \label{eq:global_energy}
    E_{\mathrm{global}}(\boldsymbol{\theta}) = \int d\mathbf{R}  \mathcal{P}(\mathbf{R}) \frac{ \langle \Psi_{\boldsymbol{\theta}}(\mathbf{R}) | H_e(\mathbf{R}^{(i)}) | \Psi_{\boldsymbol{\theta}}(\mathbf{R}) \rangle}{ \langle \Psi_{\boldsymbol{\theta}}(\mathbf{R}) | \Psi_{\boldsymbol{\theta}}(\mathbf{R}) \rangle},
\end{equation}
where the expectation value $\langle \dots \hspace{-0.001cm} \rangle $ is taken over a sufficiently large set of electronic configurations $\{\mathbf{r}\}$, with $\mathbf{r} = \{\mathbf{r}_1, \hdots, \mathbf{r}_N\}$ (assuming the same number of electrons as nuclei without loss of generality), from the conditional probability density $P_{\boldsymbol{\theta}}(\mathbf{r}|\mathbf{R}) = |\langle\Psi_{\boldsymbol{\theta}}(\mathbf{R})|\mathbf{r}\rangle|^{2} / \langle\Psi_{\boldsymbol{\theta}}(\mathbf{R})|\Psi_{\boldsymbol{\theta}}(\mathbf{R})\rangle$.
In practice, the integration over the reference probability distribution $\mathcal{P}(\mathbf{R})$ is performed through Monte Carlo averages obtained using a finite set of $M$ independent and identically distributed (i.i.d.) nuclear configurations drawn from $\mathcal{P}(\mathbf{R})$, that is, $\mathbf{R} \sim \mathcal{P}(\mathbf{R})$.
Assuming the nuclear configurations are re-sampled uniformly from an existing dataset of configurations, $\mathcal{D}$, the integral reduces to the following discrete form
\begin{equation}    \label{eq:discrete_global_energy}
    E_{\mathrm{global}}(\boldsymbol{\theta}) 
    = \frac{1}{M} \sum_{i=1}^{M} \frac{\langle \Psi_{\boldsymbol{\theta}}(\mathbf{R}^{(i)})| H_e(\mathbf{R}^{(i)}) |\Psi_{\boldsymbol{\theta}}(\mathbf{R}^{(i)})\rangle}{\langle \Psi_{\boldsymbol{\theta}}(\mathbf{R}^{(i)})|\Psi_{\boldsymbol{\theta}}(\mathbf{R}^{(i)})\rangle}
    \equiv \frac{1}{M} \sum_{i=1}^M E(\mathbf{R}^{(i)}, \boldsymbol{\theta}),
\end{equation}
where a uniform distribution $\mathcal{P}(\mathbf{R}) = 1 / M$ is implicitly used and each nuclear configuration $\mathbf{R}^{(i)}$ is part of the dataset $\mathcal{D}$, i.e. $\mathbf{R}^{(i)} \in \mathcal{D}$.
We further implicitly defined the expectation value of $H_e(\mathbf{R}^{(i)})$ with respect to the globally optimized wave function $\Psi_{\boldsymbol{\theta}}$, evaluated at the set of (global) variational parameters $\boldsymbol{\theta}$ and at one nuclear configuration $\mathbf{R}^{(i)}$, as $E(\mathbf{R}^{(i)}, \boldsymbol{\theta})$.
The global optimization strategy we propose, analogous to the framework described in the Foundation Neural-Network Quantum States paper \cite{2025_rende_foundation_neural_network_quantum_states}, leverages an extended stochastic reconfiguration (SR) scheme \cite{1998_sorella_sr,2005_sorella_sr,2017_becca_sorella_book} for multiple auxiliary systems (here each labeled by the nuclear configuration $\mathbf{R}$). 
For completeness, we rewrite this procedure, simply adapting the notation, which states that the variational parameter update $\boldsymbol{\delta \theta}$ is given by the following matrix equation
\begin{equation}
    \mathcal{S} \boldsymbol{\delta \theta} = -\eta \mathbf{F}, 
\end{equation}
where the quantum geometric tensor $\mathcal{S}$ has size $P \times P$, with $P$ the total number of variational parameters, $\eta$ is the learning rate and $\mathbf{F} = \nabla_{\boldsymbol{\theta}} E_\mathrm{global}(\boldsymbol{\theta})$ is the gradient of the loss function. 
The component of the latter $\mathcal{F}_k \in \mathbf{F}$, with $k \in {1,\hdots,P}$, are formally defined as $\mathcal{F}_k = \int d\mathbf{R} \mathcal{P}(\mathbf{R}) F_k(\mathbf{R})$, where $F_k(\mathbf{R})$ can be written as a covariance of the form
\begin{align}
    F_k(\mathbf{R}) &= -2 \mathrm{Re} \{ \langle \Delta O_k^\dagger(\mathbf{R}) \Delta H_e(\mathbf{R}) \rangle \} \nonumber \\ 
    &= -2 \mathrm{Re} \{ \langle O_k^\dagger(\mathbf{R}) H_e(\mathbf{R}) \rangle - \langle H_e(\mathbf{R}) \rangle \langle O_k^\dagger(\mathbf{R}) \rangle \},
\end{align}
where $\Delta A = A - \langle A \rangle$ corresponds to the shifted-mean operator of an arbitrary operator $A$, and $O_k(\mathbf{R}) = \partial \ln \Psi_{\boldsymbol{\theta}}(\mathbf{r},\mathbf{R})/\partial \theta_k$, with $\theta_k \in \boldsymbol{\theta}$.
The extension of the quantum geometric tensor to the extended space is performed similarly, so that $\mathcal{S}_{kl} = \int d \mathbf{R} \mathcal{P}(\mathbf{R}) S_{kl}(\mathbf{R})$, with
\begin{align}
    S_{kl}(\mathbf{R}) 
    &= \mathrm{Re} \{ \langle \Delta O_k^\dagger(\mathbf{R}) \Delta O_l(\mathbf{R}) \rangle \} \nonumber \\
    &= \mathrm{Re} \{ \langle O^\dagger_k(\mathbf{R}) O_l(\mathbf{R}) \rangle - \langle O_k^\dagger(\mathbf{R}) \rangle \langle O_l(\mathbf{R}) \rangle \}.
\end{align}
This global optimization scheme contrasts sharply with the conventional approach of individually optimizing the electronic wave function for each nuclear configuration.
Although the latter will generally provide more accurate absolute energies, it is exceedingly inefficient for large datasets and requires extensive computational resources and human intervention.

In the context of high-pressure hydrogen, we considered the Dense hydrogen DMC database \cite{2023_niu_qmc_database_hydrogen,QMC-hamm}, that we denote as $\mathcal{D}$, containing $|\mathcal{D}|=17\,528$ configurations spanning pressures from 50 to 200 GPa and temperatures from 600 to 2200 K. For preliminary testing, we restrict our attention to the 50 GPa subset $\mathcal{D}_{50}=\{x\in\mathcal{D}: P(x)=50 \ \mathrm{GPa}\}$ with $|\mathcal{D}_{50}|=2\,469$. A training set $\mathcal{C}\subset\mathcal{D}_{50}$ of size $M=512$ is selected without imposing any temperature bias, leaving $|\mathcal{D}_{50}\setminus\mathcal{C}|=1\,957$ configurations outside the learning set.
The pressure of $P = 50$ GPa was chosen as it lies farthest from the metallic regime, allowing us to test the wave function beyond the metallic phase where it performs well.
We then obtained an optimized set of parameters
$\boldsymbol{\theta}^*$ by minimizing \cref{eq:discrete_global_energy} over this subset of configurations $\mathcal{C}$.
In particular, the SR scheme described above was used to optimize all the 2889 parameters included in the trial wave function.

Using a $4\times4\times4$ twist grid (matching the protocol of Refs.~\cite{2023_niu_qmc_database_hydrogen,QMC-hamm}), the average difference between the global NQS and DMC reference energies (per electron), given by
\begin{equation}
    \Delta E = \frac{1}{M} \sum_{i=1}^M \left(\frac{E(\mathbf{R}^{(i)}, \boldsymbol{\theta}^*)}{N} - \frac{E_\mathrm{ref}(\mathbf{R}^{(i)})}{N} \right),
\end{equation}
is $\Delta E \sim 4$ mHa for configurations included in the training set, and $\Delta E \sim 5$ mHa for configurations excluded from the training set but still tagged with a DFT pressure of $P = 50$ GPa (defining a so-called test set).
(Note that we have
not corrected for two-body finite size effects \cite{2016_holzmann_finite_size_effects}
in both NQS and database DMC energies.)
As hinted in the main text, the global optimization results' accuracy is therefore not yet satisfactory, being on average more than 4 mHa away from the reference energies, leaving a lot of place for improvement.

However, in terms of the computation time, the global optimization routine ran once on a single NVIDIA H100 GPU for approximately 72 hours, with an additional 15 hours required to compute 500 test-set energies, for instance, amounting to roughly 100 GPU hours in total.
In contrast, a conventional DMC procedure, in which the electronic wave function is independently optimized for each configuration, can easily require several orders of magnitude more computational time as well as more human oversight.

\end{document}